\documentclass[pdflatex,sn-nature]{sn-jnl}

\usepackage{subcaption}
\usepackage{graphicx}%
\usepackage{caption}
\usepackage{tikz}
\usepackage{acro}
\usepackage{nameref}
\usepackage[section]{placeins}
\usepackage{float}
\usepackage{multirow}%
\usepackage{amsmath,amssymb,amsfonts}%

\usepackage{amsthm}%
\usepackage[scr=rsfs]{mathalpha}%
\usepackage[title]{appendix}%
\usepackage{acro}%
\usepackage{braket}
\usepackage{xcolor}%
\usepackage{csquotes}
\usepackage{textcomp}%
\usepackage{physics}
\usepackage{manyfoot}%
\usepackage{booktabs}%
\usepackage{algorithm}%
\usepackage{algorithmicx}%
\usepackage{algpseudocode}%
\usepackage{listings}%
\usepackage[most]{tcolorbox}
\usepackage{float}
\usepackage{siunitx} 

\theoremstyle{thmstyleone}%

\usepackage[normalem]{ulem}

\theoremstyle{thmstyletwo}%

\theoremstyle{thmstylethree}%
\newtheorem{definition}{Definition}%

\newcommand{\blue}[1]{\textcolor{blue}{#1}}

\tcbuselibrary{skins} 

\newtcolorbox{redbox}{
  colback=yellow!10,    
  colframe=red!75,      
  boxrule=0.5pt,        
  arc=4pt,              
  left=4pt,             
  right=4pt,
  top=4pt,
  bottom=4pt,
  breakable             
}

\newtcolorbox{bluebox}{
  colback=yellow!10,    
  colframe=blue!75,      
  boxrule=0.5pt,        
  arc=4pt,              
  left=4pt,             
  right=4pt,
  top=4pt,
  bottom=4pt,
  breakable             
}

\renewcommand{\blue}[1]{#1}

\newcommand{\mc}{\mathcal}

\DeclareAcronym{CSI}{
short = CSI,
long = channel state information 
}
\DeclareAcronym{FSO}{
short = FSO,
long = free-space optical
}

\DeclareAcronym{CPTP}{
short = CPTP,
long = completely positive and trace-preserving 
}
\DeclareAcronym{DV}{
short = DV,
long = discrete-variable
}

\DeclareAcronym{MIMO}{
short = MIMO,
long = multiple-input multiple-output
}

\DeclareAcronym{SDP}{
short = SDP,
long = semidefinite program
}

\DeclareAcronym{QuMIMO}{
short = QuMIMO,
long = quantum multiple-input multiple-output
}

\DeclareAcronym{QKD}{
short = QKD,
long = quantum key distribution
}

\DeclareAcronym{QEC}{
short = QEC,
long = quantum error correction
}

\begin{document}

\title{Design and \blue{Optimization} of Adaptive Diversity \blue{Schemes} in Quantum MIMO Channels}

\author*[1]{\fnm{Shehbaz} \sur{Tariq}}\email{shehbaz.tariq@uni.lu}
\equalcont{These authors contributed equally to this work.}
\author[1]{\fnm{Symeon} \sur{Chatzinotas}}\email{symeon.chatzinotas@uni.lu}
\equalcont{These authors contributed equally to this work.}

\affil*[1]{\orgdiv{Interdisciplinary Centre for
Security, Reliability, and Trust (SnT)}, \orgname{University of Luxembourg}, \orgaddress{ \postcode{L-1855}, \city{Luxembourg City}, \country{Luxembourg}}}


\abstract{
As quantum networks evolve toward a full quantum Internet, reliable transmission in \ac{QuMIMO} settings becomes essential, yet remains difficult due to noise, crosstalk, and the mixing of quantum information across subchannels. To improve reliability in such settings, we study an adaptive diversity strategy for discrete-variable QuMIMO systems based on universal asymmetric cloning at the transmitter and probabilistic purification at the receiver. An input qubit is encoded into $M$ approximate clones, transmitted over an $N\times N$ multi-mode quantum channel, and recovered through a purification map optimized using available \ac{CSI}. \blue{For the given cloning asymmetry parameters}, we derive an eigenvalue-based expression for the decoder-optimal end-to-end fidelity in the form of a generalized Rayleigh quotient, which enables efficient tuning of the cloner without iterative optimization. \blue{As a design choice, we employ \ac{SDP} to construct the purification map for only a targeted success probability $p$}. This numerical framework is used to study fixed-noise, dimension-scaling noise, and stochastic depolarization regimes. A \blue{cloning asymmetry index} is introduced to quantify the \blue{distribution of quantum information across} the multiple subchannels across these operating conditions. The results show that the proposed scheme yields significant fidelity gains in crosstalk-dominated settings and automatically adapts to channel symmetry and channel conditions. This work provides design guidelines for future QuMIMO systems and establishes a robust baseline for more advanced transmission and decoding strategies. 
}


\maketitle

\section*{Introduction}\label{sec1}
Recent years have witnessed remarkable advances in quantum information science, particularly in quantum computing \cite{KEA:23:Nat, LIS:21:Sci} and quantum communications \cite{GT:07:NP, CZC:21:Nat, P:21:PRR}. Quantum communication systems leverage the fundamental principles of superposition and entanglement to enable intrinsically secure information transfer, forming a key component of emerging quantum network infrastructures \cite{ZSHNWH:25:IEEE_O_CSTO}. These technologies are envisioned as the backbone of the quantum Internet, supporting distributed quantum computing, entanglement-assisted sensing, and long-distance quantum information exchange \cite{WEH:18:Sci, Wilde:13:book, HCMP:2025:arXiv}. Despite this progress, achieving reliable quantum-state transmission over fiber-based or free-space optical channels remains a significant challenge. Practical links suffer from photon loss, decoherence, pointing errors, and background noise \cite{SDSJM:21:IEEE_O_CSTO}, all of which directly degrade \blue{end-to-end} fidelity, limit secret-key generation rates, and constrain overall scalability \cite{PLOB:17:NC}. Interestingly, analogous limitations were encountered in classical wireless systems prior to the development of \ac{MIMO} architectures, which dramatically improved throughput, reliability, and spectral efficiency through spatial diversity and spatial multiplexing gains \cite{Fos:96:BLTJ, FGV:99:JSAC, Tel:99:ETT, ZT:03:IEEE_T_IT}. Today, \ac{MIMO} technology constitutes a core element of modern wireless standards, spanning IEEE WLAN protocols to 5G and beyond \cite{LLSAZ:14:IEEE_J_STSP}. \blue{Although \ac{MIMO} strategies were originally developed to optimize radio frequency links, their adaptation to \ac{FSO} channels has similarly yielded significant diversity and multiplexing improvements~\cite{SC:02:GLOBECOM02,razavi2005wireless, jamali2015ber, jaiswal2018free}. Such systems are typically modeled with $N_t$ optical sources and $N_r$ detection units.}

Inspired by these developments, the \ac{QuMIMO} paradigm has emerged as a pragmatic direction for improving robustness and resource utilization in quantum communication networks. Early \ac{QuMIMO} studies explored multi-mode architectures to enhance key-generation rates in \ac{QKD} systems \cite{GA:06:JLT}, while later works investigated rate–fidelity trade-offs and spatial multiplexing in multi-mode quantum channels \cite{YC:20:TCOM, KDM:21:IEEE_J_COML, KMC:23:TQE, ZPD:23:TQE}. Recent advances have extended these concepts toward quantum diversity, addressing crosstalk and depolarization through structured multi-channel encoding and joint recovery techniques \cite{WRA:25:TCOM, ROK:24:EPJ, KOuRCh:25:CommPhys, uRRK:25:IEEE_CL, TRC:2025:arXiv, KC:25:npj_QI}. In classical \ac{MIMO} systems, multiplexing gain arises from transmitting independent data streams, whereas diversity gain enhances reliability by sending redundant signals across independent fading channels. Extending these principles to the quantum domain is, however, nontrivial. The no-cloning theorem \cite{WZ:82:Nat} prohibits perfect replication of arbitrary quantum states, fundamentally constraining the realization of quantum diversity. Although this limitation does not affect classical information encoded within quantum carriers—as in \ac{QKD}—it becomes critical when the \blue{information source} itself is quantum, as in distributed quantum computing or sensing. In this regime, approximate quantum cloning offers a practical means for information spreading across multiple quantum channels, at the cost of introducing controlled distortion and correlations among the generated copies \cite{K:16:QIC, ZZZZCLKP:05:PRL, SIG:05:RMP}. Moreover, probabilistic purification techniques can be applied at the receiver to reconstruct higher-fidelity states from these degraded ensembles \cite{YCH:25:QST}. Together, these operations establish the foundation for quantum spatial diversity, enabling redundancy-driven robustness within \ac{QuMIMO} communication systems.

Building upon these foundations, this paper develops a unified end-to-end framework for adaptive diversity optimization in \ac{DV} \ac{QuMIMO} channels. The proposed model incorporates all essential stages of a practical quantum multi-channel link: asymmetric quantum cloning at the transmitter, a structured noise model consisting of independent depolarization on each subchannel together with stochastic crosstalk that redistributes quantum information across paths, and a purification stage at the receiver. This representation captures the dominant physical impairments in multi-mode quantum systems, where depolarization reflects mode-dependent decoherence and optical loss, while crosstalk accounts for coherent or incoherent mixing caused by imperfect mode separation, scattering, or hardware-induced coupling.

To characterize end-to-end performance, \blue{we exploit \ac{CSI} from the noise maps and side information about the encoder's cloning asymmetry on Haar-distributed pure states.} This statistical description enables a purification strategy that adapts to the underlying \blue{encoding strategy} and channel characteristics. Within this framework, we examine how the cloning asymmetry shapes information redundancy, inter-channel correlations, and the achievable fidelity after purification. By adjusting the asymmetry according to the available \ac{CSI}, the model provides concrete guidelines for achieving high-fidelity transmission under different noise correlations and crosstalk strengths.

The analysis considers both symmetric channels, where each subchannel experiences identical noise, and asymmetric channels with unequal depolarization levels. For each class, two operating regimes \blue{with respect to dimension scaling} are studied: one where the total noise budget is fixed and redistributed across subchannels, and another where the noise scales with the number of channels. These cases highlight how the balance between redundancy and channel heterogeneity determines the attainable diversity and fidelity. Robustness is further examined under imperfect \ac{CSI}. In many practical scenarios \blue{such as time-varying quantum channels \cite{ur2023estimating}}, instantaneous channel knowledge is unavailable and only average channel statistics are accessible. In this setting, we investigate whether probabilistic purification continues to provide improvements over direct transmission, especially when crosstalk disrupts channel ordering and degrades the usefulness of mode ranking. The results reveal how diversity, asymmetry, and inter-channel mixing jointly govern the reliability of quantum-state transfer in \ac{QuMIMO} communication systems.

\begin{figure}
    \centering
    \includegraphics[width=0.8\linewidth]{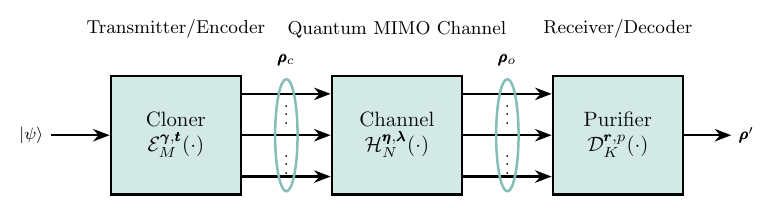}
    \caption{
A block-level representation of the \ac{QuMIMO} communication system. The encoder (transmitter) employs a cloning map $\mc{E}_M^{\pmb{\gamma},\pmb{t}}\!\left(\cdot\right)$ 
to generate multiple imperfect copies of the input state $\rho = \ketbra{\psi}$.  These copies propagate through the \ac{QuMIMO} channel 
$\mc{H}_N^{\eta,\pmb{\lambda},\delta}\!\left(\cdot\right)$, 
which models crosstalk and depolarizing noise. The decoder (receiver) applies a purification map $\mc{D}_K^{\pmb{r},p}\!\left(\cdot\right)$ 
to retrieve the quantum state $\boldsymbol{\rho}'$.
}

    \label{fig:block}
\end{figure}

\section*{Results}\label{res}
We evaluate the proposed adaptive diversity optimization framework in a single-user quantum communication setting comprising a \blue{quantum source, a transmitter/encoder}, an $N$-mode \blue{\ac{MIMO}} channel array, and a corresponding $K$-mode \blue{receiver/decoder}. The source \blue{outputs} a pure state $\boldsymbol{\rho} = \ketbra{\psi}{\psi}$, which is encoded into $M$ imperfect clones using the asymmetric cloning map $\mathcal{E}_{M}^{\pmb{\gamma},\pmb{t}}(\cdot)$, where $\pmb{\gamma}$ is the vector of cloning asymmetry coefficients and $\pmb{t}$ specifies the subchannels selected for transmission. The asymmetry vector $\pmb{\gamma}$ determines the encoded multi-clone state $\boldsymbol{\rho}_{\mathrm{c}}$, whose local marginals $\boldsymbol{\rho}_{\mathrm{c},k}$ for $k\in\{1,\ldots,M\}$ achieve the target fidelities $\{F_k\}$ and define the operational asymmetry point of the cloning process. After the cloning stage, the encoded state 
$\boldsymbol{\rho}_{\mathrm{c}}$ 
is transmitted through an $N\times N$ quantum MIMO channel 
$\mathcal{H}_{N}^{\eta,\boldsymbol{\lambda}, \delta}(\cdot)$, 
which jointly models intra-channel depolarization and inter-channel 
crosstalk impairments. The channel is formulated as a doubly stochastic 
composition of two \ac{CPTP} maps,
\begin{align}
\mathcal{H}_{N}^{\eta,\boldsymbol{\lambda}, \delta}
   = \mathcal{N}_{\boldsymbol{\lambda}}\!\circ\!\mathcal{X}_{\eta, \delta},
\end{align}
where $\mathcal{N}_{\boldsymbol{\lambda}}$ describes independent 
per-branch depolarization and $\mathcal{X}_{\eta,\delta}$ models global 
crosstalk through a stochastic ensemble of permutation unitaries, with 
$\boldsymbol{\lambda}$ the depolarization vector, $\eta$ the crosstalk 
strength, and $\delta$ the spatial decay parameter.
The doubly stochastic structure ensures complete positivity, trace 
preservation, and unitality of the composite channel~\cite{W:09:QIC}.

\begin{definition}[Local depolarization]
Each subchannel $i$ is represented by a single-qubit depolarizing map 
with Kraus operators $\{K^{(i)}_k\}$ satisfying 
$\sum_k {K^{(i)}_k}^{\!\dagger}K^{(i)}_k = I$. 
The overall depolarizing map $\mathcal{N}_{\boldsymbol{\lambda}}$ is 
constructed as the tensor product of all local maps,
\begin{align}
   A_k
   = K^{(1)}_{k_1}\!\otimes\!K^{(2)}_{k_2}\!\otimes\!\cdots\!\otimes\!K^{(N)}_{k_N},
   \qquad k=(k_1,\ldots,k_N),
\end{align}
so that each $A_k$ acts collectively on all $N$ subchannels and forms a 
Kraus operator of the base noise ensemble prior to crosstalk.
\end{definition}
\begin{figure}
    \centering
    \includegraphics[width=1\linewidth]{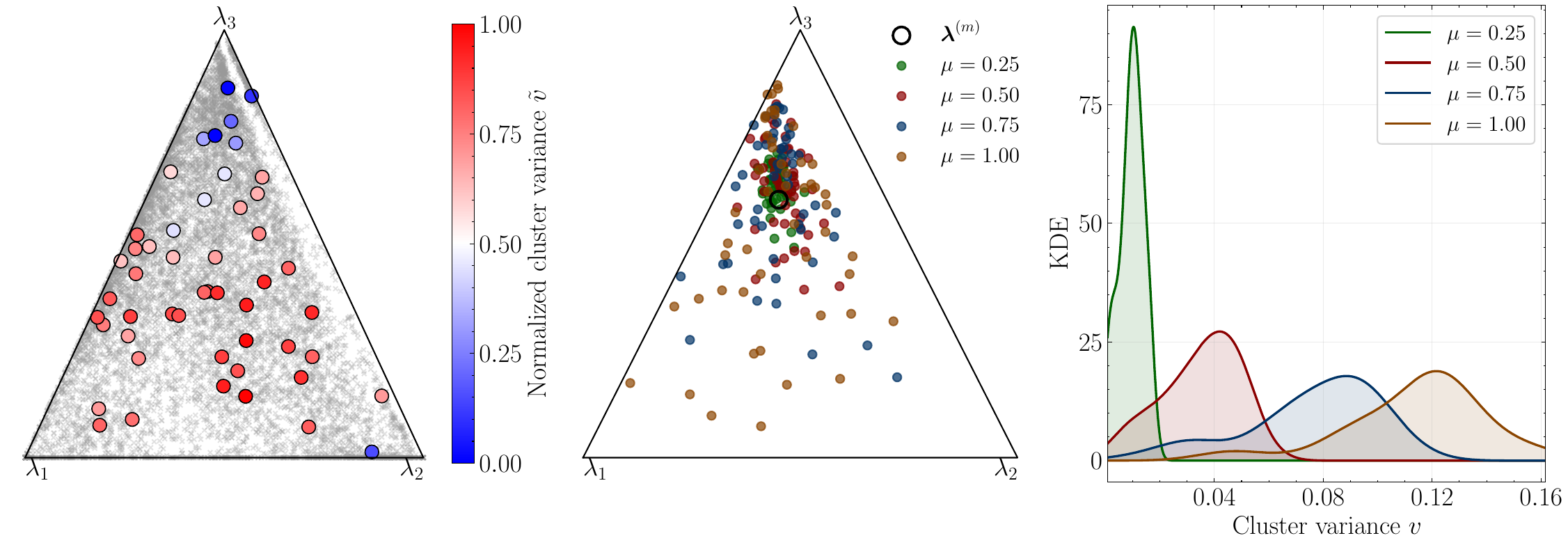}
   \caption{Sampling illustration for a three–channel configuration ($N=3$) with total
depolarization budget $Z=1.0$.  
\textbf{Left:} Ternary plot of the $50$ mean depolarization allocations 
$\boldsymbol{\lambda}^{(m)}$ sampled from $\mathcal{A}_Z$.  
\textbf{Middle:} For a selected mean vector, ternary plots of the $50$ perturbed 
realizations $\mathbf{x}^{(m)}(\mu)$ generated for each noise level 
$\mu\in\{0.25,0.50,0.75,1.00\}$, illustrating how dispersion around the mean increases  
with~$\mu$.  
\textbf{Right:} Kernel density estimates (KDEs) of the absolute cluster variance $v$, 
computed over the perturbations generated for each mean vector and each noise level, 
showing the shift toward higher variability as~$\mu$ increases.}

    \label{fig:ternary}
\end{figure}

Moreover, as the dimension of the system increases, each additional subchannel introduces an
independent depolarization parameter, which causes the noise landscape to expand
exponentially with $N$. Exhaustively evaluating all such configurations
is infeasible, so we adopt a sampling-based stochastic noise model that
captures representative behaviour without exploring the full
high-dimensional space. Moreover, under rapidly varying conditions,
instantaneous CSI is not available; therefore, the system relies on
first-order statistical information on depolarization noise levels
$\boldsymbol{\lambda}^{(m)}$, where each mean allocation satisfies
$\sum_{i=1}^N \lambda_i^{(m)} = Z$ for a fixed total depolarization
budget $Z>0$. These mean allocations are drawn from the simplex of feasible
total-noise budgets and are subsequently perturbed by multiplicative
Gamma--Gamma fluctuations whose strength is controlled by a parameter
$\mu>0$. Small $\mu$ produces realizations close to the mean profile,
while larger $\mu$ induces stronger deviations. This provides a scalable
and statistically meaningful model of depolarizing noise in the \ac{QuMIMO} system, as shown in Fig.~\ref{fig:ternary}.

\begin{figure}
    \centering
    \includegraphics[width=1\linewidth]{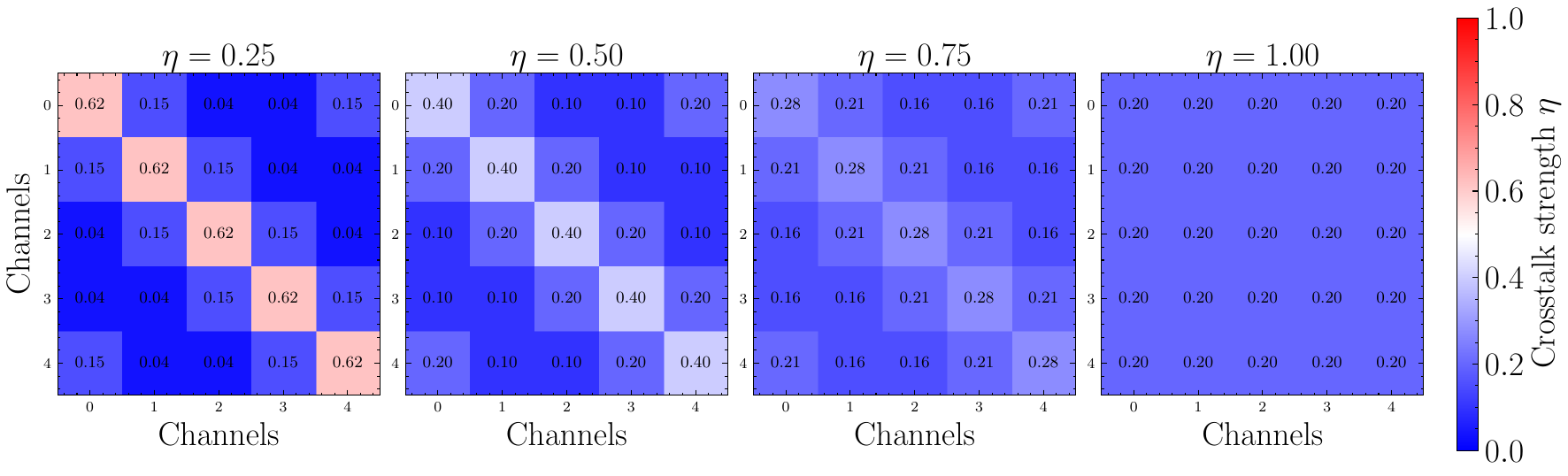}
    \caption{
Stochastic coupling matrices $P(\eta,\delta)$ for $N=5$ channels, 
illustrating the spatial coupling pattern under different crosstalk strengths 
$\eta = \{0.25,\,0.5,\,0.75,\,1.0\}$ with fixed decay exponent $\delta = 1$. 
As $\eta$ increases, the off-diagonal coupling elements become more pronounced, 
indicating stronger inter-channel mixing.
}
    \label{fig:crosstalk}
\end{figure}

\begin{definition}[Stochastic crosstalk model]
In \ac{DV} quantum information processing, spatially distributed 
subchannels may partially mix, causing information in one path to leak 
into others and thereby \blue{reducing end-to-end fidelity}. 
This inter-channel mixing is described using a stochastic unitary 
process generated by permutation operators, which reorder channel 
indices while preserving global unitarity and the trace of the system. 
This representation captures realistic effects in optical and trapped-ion 
architectures, where correlated interference or mode mixing arises from 
imperfect spatial isolation~\cite{RSLLRSMMBSO:25:arXiv,LTDW:25:PRA}. Spatial selectivity is controlled through a decay parameter $\delta>0$, 
leading to the pairwise coupling kernel
\begin{align}
   c_{ij}
   &= \frac{e^{-\delta\, d(i,j)}}
           {\sum_{k=1}^{N} e^{-\delta\, d(i,k)}},
   &
   d(i,j) &= \min(|i-j|,\, N-|i-j|),
\end{align}
which yields the stochastic matrix $P(\eta,\delta)$ governing local 
mixing probabilities. The parameter $\eta\in[0,1]$ sets the overall 
crosstalk strength. Each permutation $\pi\in S_N$ corresponds to a unitary
\begin{align}
   U_\pi\ket{a_1,a_2,\ldots,a_N}
   = \ket{a_{\pi^{-1}(1)},a_{\pi^{-1}(2)},\ldots,a_{\pi^{-1}(N)}},
\end{align}
with the relations $U_\pi U_\sigma = U_{\pi\circ\sigma}$ and 
$U_\pi^{\dagger} = U_{\pi^{-1}}$. Global permutation weights are
\begin{align}
   c_\pi
   &= \frac{\prod_{i=1}^{N} c_{i,\pi(i)}}
           {\sum_{\pi'\in S_N} 
             \prod_{i=1}^{N} c_{i,\pi'(i)}},
   &
   \sum_{\pi\in S_N} c_\pi &= 1,
\end{align}
ensuring consistency with the local coupling kernel. The resulting crosstalk channel is
\begin{align}
   \mathcal{X}_{\eta,\delta}(\boldsymbol{\rho})
   = (1-\eta)\,\boldsymbol{\rho}
     + \eta \sum_{\pi\in S_N} c_\pi\,
       U_\pi\,\boldsymbol{\rho}\,U_\pi^{\dagger},
\end{align}
interpolating between no mixing ($\eta=0$) and uniform random mixing 
($\eta=1$). 
\end{definition}
Merging crosstalk with local depolarization yields the complete channel
\begin{align}
   \mathcal{H}_N^{\eta,\boldsymbol{\lambda},\delta}(\boldsymbol{\rho})
   = (1-\eta)\sum_{k}A_k\,\boldsymbol{\rho}\,A_k^{\dagger}
     + \eta\!\!\sum_{\pi\in S_N}\! c_\pi
       \sum_{k} U_\pi A_k\,\boldsymbol{\rho}\,A_k^{\dagger} U_\pi^{\dagger},
\end{align}
which is completely positive and trace-preserving. 
For weak coupling ($\eta\!\ll\!1$), the channel approximates 
independent depolarized branches; for strong coupling ($\eta\!\to\!1$), 
the outputs become fully mixed. The propagated state at the receiver is
\begin{align}
   \boldsymbol{\rho}_{\mathrm{out}}
   = \mathcal{H}_N^{\eta,\boldsymbol{\lambda},\delta}
     \!\left(\boldsymbol{\rho}_{\mathrm{c}}\right),
\end{align}
which serves as the input to the purification-based decoding stage. The receiver obtains this noisy joint output state $\boldsymbol{\rho}_{\mathrm{out}}$ and performs post-processing to estimate $\ket{\psi}$. The overarching objective is to maximize \blue{average fidelity, exploiting knowledge of the \ac{QuMIMO} channel.} The Choi maps of the cloning and channel operations, $\boldsymbol{J}^{\mathcal{E}}$ and $\boldsymbol{J}^{\mathcal{H}}$, jointly determine the \blue{side information} used for adaptive optimization.

At the receiver, the $N$ output modes are processed by the probabilistic purification map $\mathcal{D}_{K}^{\pmb{r},p}(\cdot)$ to obtain a high-fidelity estimate of the transmitted state, where $\pmb{r}$ specifies the $K$ subchannels selected for decoding and $p$ is the purification success probability following the framework of~\cite{YCH:25:QST}. The cloning asymmetry vector $\pmb{\gamma}$ is optimized using available \ac{CSI} so that the transmitted clones exploit the diversity structure of the channel, thereby maximizing the end-to-end fidelity. The framework is adaptive, operating effectively across symmetric depolarizing channels, asymmetric configurations with heterogeneous noise levels, and a range of crosstalk strengths. The detailed formulations of the universal asymmetric cloner, the probabilistic purification decoder, and the proposed asymmetry optimization technique are provided in the \nameref{Methods} section. This clone–distribution–channel–purification pipeline forms the mathematical backbone of the fidelity analysis developed in the subsequent sections.

\subsubsection*{Operating Regimes}
Building on the stochastic noise model introduced earlier, we investigate the
average fidelity gain achieved by the adaptive communication strategies proposed across a set of representative operating regimes. For each regime, we evaluate performance under
both \emph{symmetric} depolarizing channels—where all subchannels experience the same
noise level—and \emph{asymmetric} configurations in which each subchannel is subject to a
different depolarizing impairment. This allows us to assess the robustness of the proposed
framework under homogeneous, heterogeneous, and statistically fluctuating channel
conditions. 
\begin{definition}[End-to-end fidelity]
    For a pure input state $\ket{\psi}$ and a fixed encoder--channel--decoder
cascade, the end-to-end fidelity is defined as
\begin{align}
   \mathcal{F}(\psi)
   = \bra{\psi}\,\hat{\rho}(\psi)\,\ket{\psi},
   \label{eq:single_state_fidelity}
\end{align}
where $\hat{\rho}(\psi)$ denotes the retrieved single-qubit output
obtained by applying the probabilistic decoder
$\mathcal{D}_{K}^{\boldsymbol{r},p}$ to the noisy post-channel state.
Let
\begin{align}
   \rho_A(\psi)
   = \Lambda_A\!\bigl(\rho\bigr)
   = \mathcal{H}_{N}^{\eta,\boldsymbol{\lambda},\delta}
     \!\bigl(
        \mathcal{E}_M^{\boldsymbol{\gamma},\boldsymbol{t}}(\rho)
      \bigr),
   \qquad
   \rho = \ket{\psi}\!\bra{\psi},
\end{align}

denote the state of $K$ bit at the decoder input (obtained by selecting
$K$ modes from the $N$-mode output of the channel, as specified by
$\boldsymbol{r}$), where $\Lambda_A$ is the effective end-to-end map from the transmitor to the receiver.  For a given success probability $p$, the CPTNI map
$\mathcal{D}_{K}^{\boldsymbol{r},p}$ is represented by its Choi operator
$\boldsymbol{J}^{\mathcal{D}}_{A\rightarrow B}$, and the purified output is
\begin{align}
   \rho_{\mathrm{p}}(\psi)
   = \mathcal{D}_{K}^{\boldsymbol{r},p}\bigl(\rho_A(\psi)\bigr)
   = \Tr_{A}\!\left[
       \boldsymbol{J}^{\mathcal{D}}_{A\rightarrow B}
       \bigl(\rho_A(\psi)^{\mathsf{T}}\!\otimes I_B\bigr)
     \right].
\end{align}
Incorporating failures as maximally mixed outcomes, the average
single-shot output state is
\begin{align}
   \hat{\rho}(\psi)
   = p\,\rho_{\mathrm{p}}(\psi) + (1-p)\,\tfrac{I}{2}.
\end{align}

The performance metric used throughout this work is the Haar-averaged
fidelity
\begin{align}
   \overline{\mathcal{F}}
   = \int \mathcal{F}(\psi)\,d\psi
   = \int 
      \bra{\psi}\,\hat{\rho}(\psi)\,\ket{\psi}\,d\psi,
   \label{eq:average_fidelity}
\end{align}
which quantifies the expected input--output similarity across all pure
inputs.
\end{definition}
\blue{To benchmark performance, we evaluate the average end-to-end fidelity of the received
state over Haar-distributed pure inputs, sent through the \ac{QuMIMO} channel with stochastic crosstalk and independent depolarization on each subchannel. To allow a uniform comparison, each transmission strategy is indexed by
$s\in\{\mathrm{dir},\mathrm{pur},\mathrm{div},\mathrm{sym},\mathrm{blind}\}$,
with corresponding end-to-end average fidelity $\overline{\mathcal{F}}_{s}$.
For every strategy, the end-to-end processing chain is written as
$(\mathcal{E}^{(s)},\mathcal{H}_{N}^{\eta,\boldsymbol{\lambda},\delta},
\mathcal{D}^{(s)})$, since the channel
$\mathcal{H}_{N}^{\eta,\boldsymbol{\lambda},\delta}$ is fixed and common
to all schemes.  
Every receiver is modeled as a probabilistic purification decoder
$\mathcal{D}_{K}^{\boldsymbol{r},p}$ acting on a $K$-qubit noisy input and
producing the output state
$\tilde{\rho}=\mathcal{D}_{K}^{\boldsymbol{r},p}(\rho_{\mathrm{in}})$.
To maintain a consistent $K$-qubit decoding interface across all
strategies, the received state is always embedded in the full $N$-mode
Hilbert space expected by the channel and decoder. In schemes where
only a single subchannel is used for transmission (e.g., direct transmission with
$M{=}1$), the unused subchannels are treated as maximally mixed states
$I/2$, ensuring that the decoder $\mathcal{D}_{K}^{\boldsymbol{r},p}$
always acts on a well-defined $2^{K}$-dimensional subsystem within the
$2^{N}$-dimensional joint space. This formulation allows all strategies—direct transmission,
purification-only, symmetric or adaptive cloning, and blind
diversity—to be embedded within the same \ac{SDP}-based decoding framework. We distinguish the following strategies:}
\begin{enumerate}
    \item \textbf{Direct deterministic transmission} $(s=\mathrm{dir})$:  
    No cloning is employed ($M=1$).  The decoder is fixed to $\mathcal{D}^{(\mathrm{dir})}
\equiv\mathcal{D}_{K}^{\boldsymbol{r},p=1}$, which is \ac{CPTP} and thus
reduces to the identity channel on the information-carrying mode.  
Transmission is performed over a single transmit–receive mode pair
$(t^{\star}, r^{\star})$, where $r^{\star}$ is chosen (using CSI) as the
least depolarized receive subchannel and $t^{\star}$ is the corresponding
transmit mode mapped to $r^{\star}$ by
$\mathcal{H}_{N}^{\eta,\boldsymbol{\lambda},\delta}$, while all remaining
modes are treated as maximally mixed.

    \item \textbf{Direct probabilistic purification} $(s=\mathrm{pur})$:  
    Again $M=1$, but with $0<p<1$.  
    The SDP optimizes a $K$-mode probabilistic purifier 
    $\mathcal{D}_{K}^{\boldsymbol{r},p}$ acting on a state where only one 
    mode carries the signal, and the others are maximally mixed, thereby 
    isolating the benefit of post-selected purification without cloning-based 
    diversity.

    \item \textbf{Adaptive cloning–purification diversity} $(s=\mathrm{div})$:  
    Here $M\ge2$ clones are transmitted over modes $\boldsymbol{t}$ and 
    jointly processed by a $K$-mode decoder 
    $\mathcal{D}_{K}^{\boldsymbol{r},p}$.  
    The cloning asymmetry vector $\boldsymbol{\gamma}$, the mode selection 
    $(\boldsymbol{t},\boldsymbol{r})$, and the decoder are jointly optimized 
    using CSI, realizing a fully adaptive diversity scheme.

    \item \textbf{Symmetric cloning diversity} $(s=\mathrm{sym})$:  
    CSI is available, but the cloning asymmetry is constrained to the 
    uniform configuration $\gamma_k = 1/M$.  
    The encoder still adapts $(\boldsymbol{t},\boldsymbol{r})$ to CSI, 
    isolating the gain due solely to channel-aware mode selection under 
    symmetric cloning.

    \item \textbf{Blind cloning–purification diversity} $(s=\mathrm{blind})$:  
    No CSI is available.  
    The encoder employs the symmetric cloner $\gamma_k = 1/M$, and the 
    decoder $\mathcal{D}_{K}^{\boldsymbol{r},p}$ is designed from the 
    Haar-averaged operators $(\boldsymbol{Q},\boldsymbol{R})$ constructed 
    under an identity-channel prior, providing a non-adaptive benchmark for 
    ensemble-averaged \ac{QuMIMO} performance.
\end{enumerate}

 The direct-transmission strategy yields
$\mathcal{F}_{\mathrm{pur}}(p)$, obtained by sending the state through the least
depolarized subchannel (known through CSI) and subsequently applying purification with
success probability~$p$. The special case $\mathcal{F}_{\mathrm{dir}}$ corresponds
to deterministic decoding in which the receiver merely selects the best channel and no
probabilistic purification is performed. The adaptive asymmetric-cloning approach achieves
$\mathcal{F}_{\mathrm{div}}(p)$, where the asymmetry vector $\boldsymbol{\gamma}$ is
optimized using CSI prior to transmission through the \ac{QuMIMO} link. The fidelity under
symmetric cloning is denoted $\mathcal{F}_{\mathrm{sym}}(p)$, corresponding to the uniform
distribution of information across all branches. fidelity $\mathcal{F}_{\mathrm{blind}}(p)$
refers to symmetric cloning performed without CSI, where the receiver assumes an
identity channel when applying
purification to the outputs of the channel \ac{QuMIMO}. 
\begin{definition}[Cloning asymmetry index]
To quantify how uniformly the quantum information is distributed across
the available branches, and to provide a principled criterion for
comparing asymmetric cloning strategies, we introduce the
\emph{cloning asymmetry index} inspired by the classical Jain index
\cite{JCH:84:ERL_DEC}. The metric captures whether a strategy
concentrates clone fidelity on a few strong subchannels or spreads it
more evenly across the system. Let $\mathbf{F} = (F_1,\ldots,F_N)$ denote the vector of single-clone
fidelities, and let $F_{\mathrm{MM}}$ be the fidelity of the maximally
mixed state (for qubits, $F_{\mathrm{MM}} = \tfrac12$). We define the
effective contributions
\begin{align}
    \tilde F_i =
    \frac{F_i - F_{\mathrm{MM}}}{1 - F_{\mathrm{MM}}}\, F_i,
\end{align}
which suppress fidelities at or below the maximally mixed baseline.
The \emph{cloning asymmetry index} is then
\begin{align}
    J(\mathbf{F})
    = \frac{\bigl(\sum_{i=1}^{N} \tilde F_i\bigr)^2}
           {N \sum_{i=1}^{N} \tilde F_i^{\,2}}.
\end{align}

The index satisfies $J(\mathbf{F}) \in [1/N,\,1]$. The lower extreme
$J = 1/N$ corresponds to a highly concentrated allocation in which
effectively only a single clone contributes useful fidelity (direct
transmission). The upper extreme $J = 1$ is achieved when all branches
contribute equally, corresponding to symmetric cloning.
\end{definition}

\begin{definition}[Empirical asymmetry density]
Given the \emph{cloning asymmetry index} $J(\boldsymbol{F})$ and a set of
$L$ channel realizations with mean depolarization vectors
$\{\boldsymbol{\lambda}^{(m)}\}_{m=1}^{L}$, let
$\boldsymbol{F}^{(m)}$ denote the optimized clone fidelity vector for
realization $m$ and
\begin{align}
  J^{(m)} = J\bigl(\boldsymbol{F}^{(m)}\bigr)
\end{align}
the corresponding asymmetry index.  
The empirical density of $J$ is defined as
\begin{align}
  \widehat{\varphi}_{J}(x)
  = \frac{1}{L}\sum_{m=1}^{L}
   \zeta\!\bigl(x - J^{(m)}\bigr),
\end{align}
where $\zeta(\cdot)$ denotes the Dirac delta distribution.  
In an $N\times N$ \ac{QuMIMO} channel, if $\widehat{\varphi}_{J}(x)$
concentrates near the lower bound $1/N$, this indicates that the optimizer
effectively routes useful fidelity through a single least–noisy subchannel.
Conversely, if $\widehat{\varphi}_{J}(x)$ remains concentrated near $x\simeq 1$,
This confirms that cloning continues to distribute quantum information more
evenly across subchannels.
\end{definition}

\begin{figure}
    \centering
    \includegraphics[width=1\linewidth]{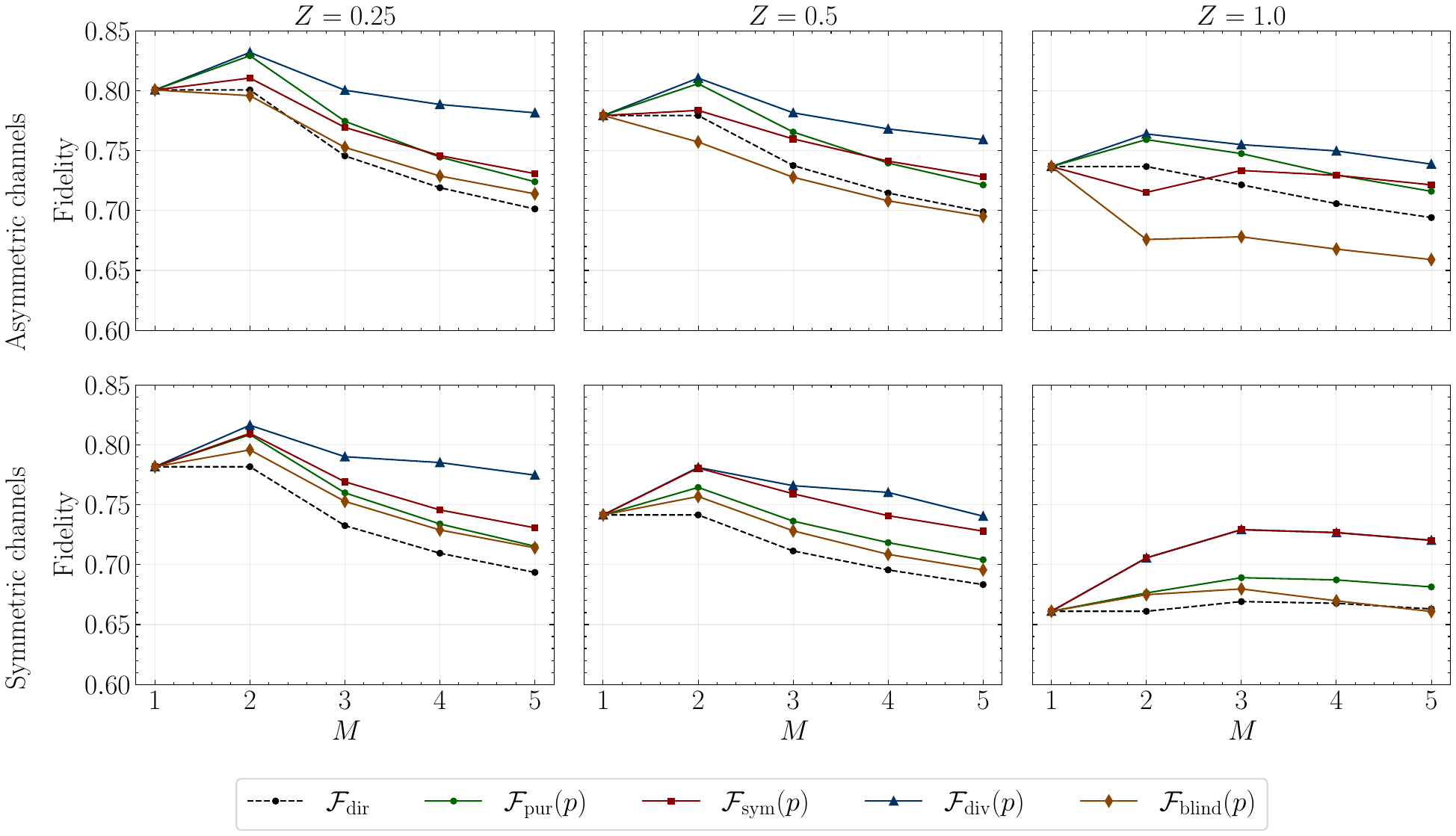}
\caption{
Fidelity versus the number of clones $M$ for an $N \times N$ \ac{QuMIMO} channel with $M=N$, under a fixed total depolarization budget $Z$. 
For each value of $Z$, the system dimension is increased by generating $M$ asymmetric clones and transmitting them across $M$ parallel subchannels, each subject to independent depolarization and stochastic crosstalk. 
Throughout the figure, the crosstalk strength is set to $\eta=0.8$ and the purification probability is fixed at $p=0.8$. 
Every curve represents the average fidelity computed over $50$ mean depolarization vectors $\boldsymbol{\lambda}^{(m)}$, ensuring a statistically stable estimate of end-to-end performance. 
The upper row corresponds to asymmetric channels, while the lower row corresponds to symmetric channels. 
Each subplot reports the end-to-end average fidelity of schemes analyzed in this work: the direct baseline $\mathcal{F}_{\mathrm{dir}}$, direct transmission at $p=0.8$, $\mathcal{F}_{\mathrm{pur}}(p)$, symmetric combining $\mathcal{F}_{\mathrm{sym}}(p)$, the blind purification benchmark $\mathcal{F}_{\mathrm{blind}}(p)$, and the proposed cloning--purification strategy $\mathcal{F}_{\mathrm{div}}(p)$. 
The results show that the optimal choice of $M$ depends on the underlying channel structure, yet the adaptive cloning--purification framework consistently achieves the highest fidelity across all configurations, demonstrating robustness to both channel symmetry and subchannel crosstalk.
}

    \label{fig:fid_Z}
\end{figure}

\subsubsection*{Fixed Level of System Noise}

In this regime, the total depolarization budget $Z$ is held fixed as the number of available branches increases. Each additional branch therefore receives a smaller share of the overall noise, yielding progressively lower per-branch depolarization with increasing system dimension. \blue{The rationale of simulating such a regime is that when depolarization arises from stray/background photons in a \ac{FSO} array, keeping the total aperture area and field of view fixed while increasing the number of array elements does not change the total background flux impinging on the receiver; the additional elements merely partition a fixed photon budget across more modes, so the overall system noise remains approximately constant in line with diffraction-limited background models for extended sources \cite{kaushal2016optical}}. Moreover, the number of admissible crosstalk operations grows, enhancing mode mixing across subchannels. These controlled-SWAP permutations do not introduce additional noise; they simply redirect quantum information between modes, all of which remain available for recovery during purification.

Fig.~\ref{fig:fid_Z} shows that, under these conditions, the proposed adaptive diversity framework consistently achieves the highest end–to–end fidelity across all channel configurations. As $N$ grows, the optimized asymmetric cloning strategy naturally concentrates weight on the most favorable subchannels, outperforming direct transmission, symmetric cloning, and blind purification. Notably, purification remains advantageous even beyond the intrinsic distortion introduced by cloning: redistributed mode components \blue{after direct transmission} can be coherently reintegrated during the purification stage, extending the effectiveness of diversity-based transmission.

\begin{figure}
    \centering
    \includegraphics[width=1\linewidth]{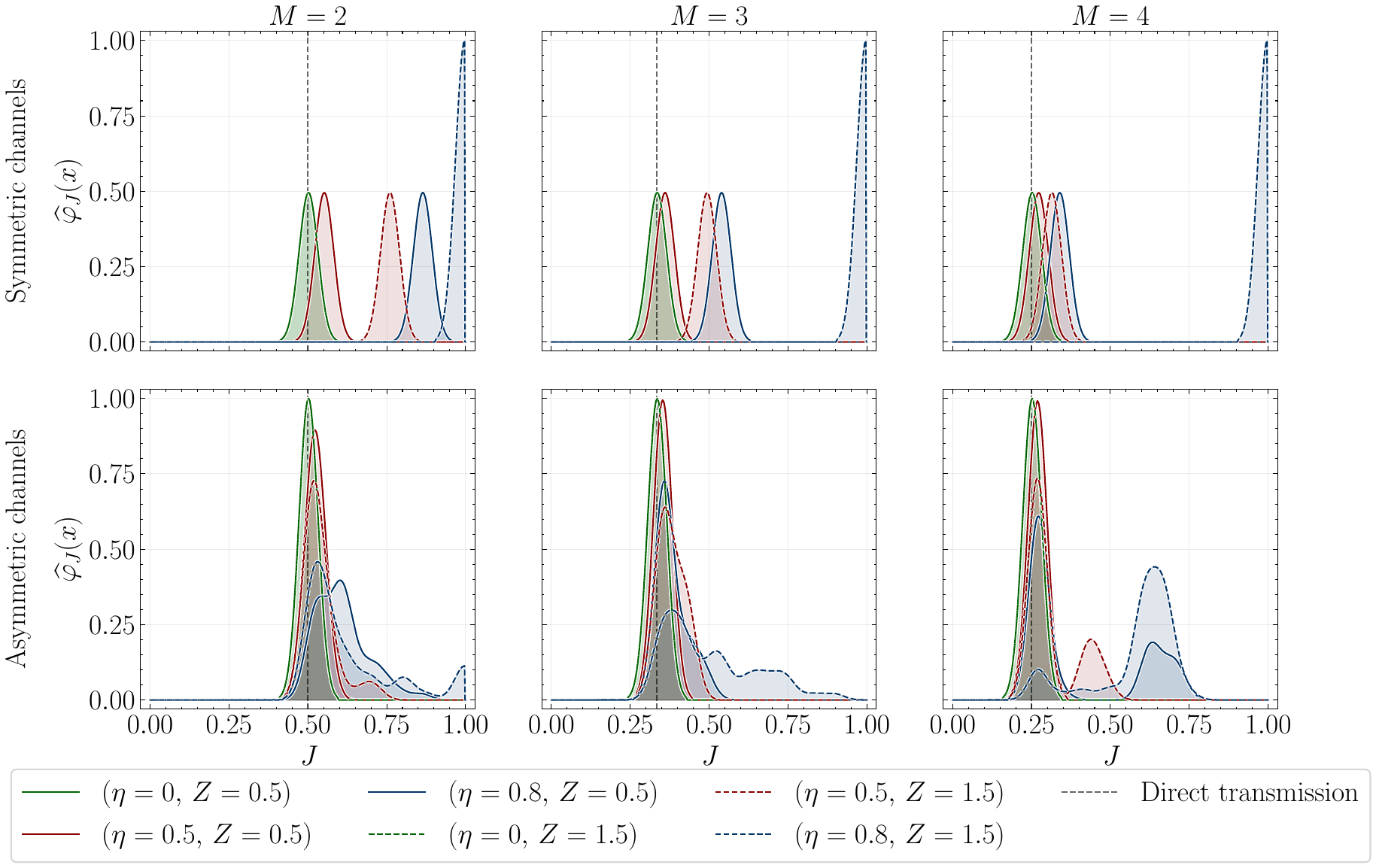}
    \caption{
Estimated probability density of the \blue{cloning asymmetry index} $J$ under a fixed depolarization budget $Z$, computed for the \emph{optimal} cloning asymmetry selected for each channel condition. 
Each panel shows the density $\widehat{\varphi}_{J}(x)$ of the  \blue{cloning asymmetry index} $J$ at $p=0.8$, averaged over $50$ mean depolarization vectors $\boldsymbol{\lambda}^{(m)}$. 
Columns correspond to $M=2,3,4$, with symmetric channels in the top row and asymmetric channels in the bottom row. 
Different crosstalk levels $\eta \in \{0, 0.5, 0.8\}$ are color-coded, and solid versus dashed curves represent distinct noise-budget settings $Z$. 
The vertical dashed line indicates the \blue{cloning asymmetry index} of direct transmission, $J = 1/M$. 
These distributions reveal how the optimal cloning asymmetry balances redundancy and channel inequality, highlighting the strategies favored under each combination of noise, symmetry, and crosstalk.
}

    \label{fig:pdf_Z}
\end{figure}

The \blue{cloning asymmetry index} distributions in Fig.~\ref{fig:pdf_Z} further illustrate how the optimization adjusts to the noise structure. In asymmetric channels, the optimal allocation progressively favors the least noisy branch, while in symmetric channels the allocation remains balanced due to the near-uniform effective noise induced by strong crosstalk. Across all settings, the observed \blue{cloning asymmetry index} relationship follows a consistent pattern: the largest diversity gains emerge in regimes dominated by mode mixing rather than pure depolarization, since swapped components are preserved and subsequently recombined. These results demonstrate the robustness of the proposed framework and its ability to select the appropriate operating strategy—cloning, purification, or direct transmission—based solely on the available \ac{CSI}.

\begin{figure}
    \centering
    \includegraphics[width=1\linewidth]{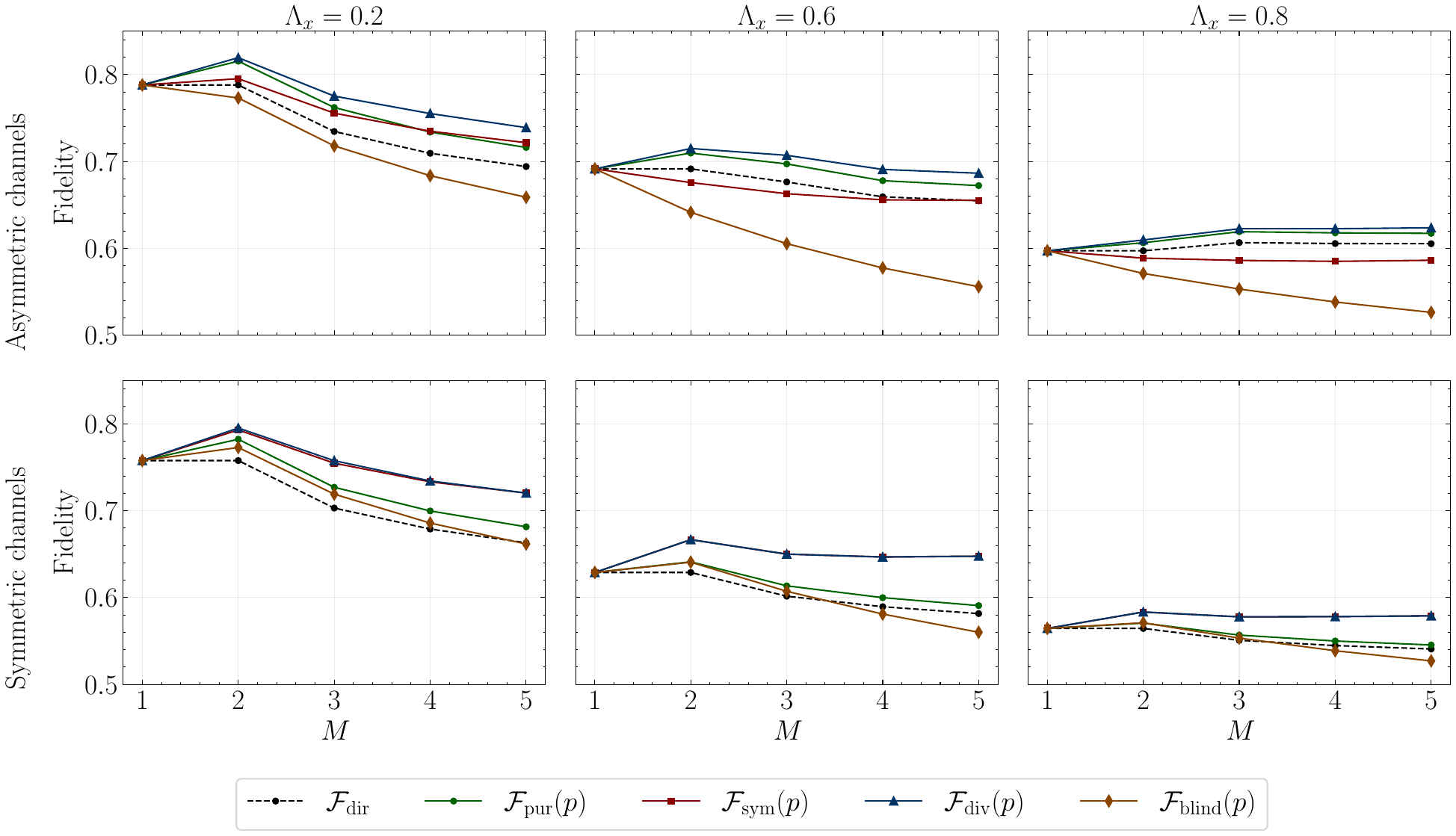}
    \caption{
Fidelity versus the number of clones $M$ for an $N \times N$ \ac{QuMIMO} channel with $M=N$, when the total depolarization budget scales with the system dimension as $Z = M \Lambda_x$. 
The columns correspond to different per-channel noise levels $\Lambda_x$, while the upper and lower rows show asymmetric and symmetric channels, respectively. 
For each setting, $M$ asymmetric clones are transmitted across $M$ subchannels subject to depolarization and crosstalk with fixed $\eta = 0.8$ and purification success probability $p = 0.8$, and the fidelities are averaged over $50$ mean depolarization vectors $\boldsymbol{\lambda}^{(m)}$. 
Each subplot compares the direct baseline $\mathcal{F}_{\mathrm{dir}}$, direct transmission at $p=0.8$, $\mathcal{F}_{\mathrm{pur}}(p)$, symmetric combining $\mathcal{F}_{\mathrm{sym}}(p)$, blind purification $\mathcal{F}_{\mathrm{blind}}(p)$, and the proposed cloning--purification strategy $\mathcal{F}_{\mathrm{div}}(p)$. 
The results show that when noise scales with $M$, the benefit of optimizing the cloning asymmetry diminishes: in asymmetric channels it is best to transmit through the least noisy subchannel, while in symmetric channels symmetric cloning remains effective up to $M=5$ due to the near-identical channel conditions induced by strong crosstalk.}

    \label{fig:fid_x}
\end{figure}

\subsubsection*{System Noise Scaling with Size}

In this regime, each additional branch contributes an equal amount of depolarizing noise, so the total noise level grows proportionally with the system dimension according to $Z = M \Lambda_x$. \blue{The motivation for this regime is that, in \ac{FSO} arrays, increasing the effective aperture area to host more spatial modes inevitably collects more background photons, so the aggregate depolarization budget naturally grows with the system size.
} This setting maintains the average fidelity of the direct transmission link while stressing the diversity mechanisms of the \ac{QuMIMO} architecture. As the system scales, both the number of noisy components and the likelihood of crosstalk increase, making this scenario particularly sensitive to the trade-off among \blue{information distribution}, distortion and \blue{retrieval fidelity}.

Fig.~\ref{fig:fid_x} shows that when noise scales with $M$, the advantages of optimizing the cloning asymmetry diminish. In asymmetric channels, the additional depolarization introduced with increasing system size overwhelms the potential benefits of distributing the quantum state across multiple branches. Consequently, the optimal strategy effectively
collapses to using the single least–noisy subchannel, and the cloning
stage offers no further gain. Importantly, purification still provides a
performance advantage: even in this degenerate single-subchannel regime,
reducing the success probability $p$ allows the decoder to post-select
cleaner realizations, yielding higher fidelity than the deterministic
($p{=}1$) case. In symmetric channels, however, cloning remains effective up to moderate dimensions ($M \le 5$), largely because strong crosstalk renders the branches nearly identical and allows purification to compensate for mode mixing. 

\begin{figure}
    \centering
    \includegraphics[width=1\linewidth]{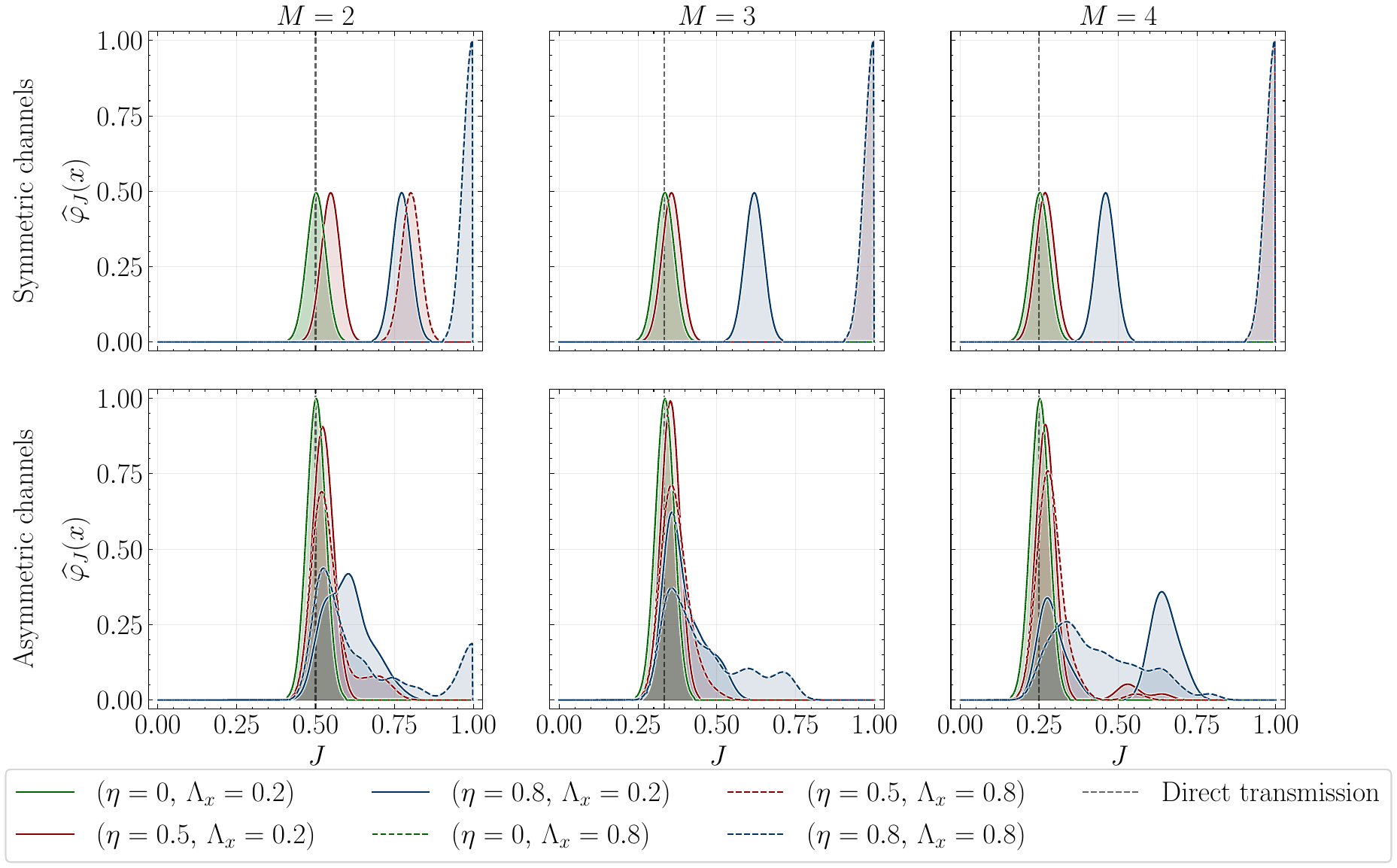}
    \caption{
Estimated probability density of the \blue{cloning asymmetry index} $J$ for the optimal cloning asymmetry when the depolarization budget scales with the system dimension as $Z = M \Lambda_x$. 
Columns correspond to $M=2,3,4$, with symmetric channels in the top row and asymmetric channels in the bottom row. 
For each $M$, curves show the density $\widehat{\varphi}_{J}(x)$ of $J$ at $p=0.8$ for different crosstalk strengths $\eta \in \{0, 0.5, 0.8\}$ and per-channel noise levels $\Lambda_x \in \{0.2, 0.8\}$, and are averaged over $50$ mean depolarization vectors $\boldsymbol{\lambda}^{(m)}$. 
The vertical dashed line marks the \blue{cloning asymmetry index} of direct transmission, $J = 1/M$. 
The distributions highlight how increasing $M$ under a scaling noise budget pushes the optimal strategy toward concentrating weight on the best subchannels in the asymmetric case, while in the symmetric case the \blue{cloning asymmetry index} remains comparatively high due to nearly identical effective channels under strong crosstalk.
}

    \label{fig:pdf}
\end{figure}

The \blue{cloning asymmetry index} distributions in Fig.~\ref{fig:pdf} reinforce these observations. As $M$ increases, the optimal allocation for asymmetric channels becomes increasingly concentrated on a single branch, reflecting the dominance of the least noisy path when overall noise grows with system size. In symmetric channels, the \blue{cloning asymmetry index} remains comparatively high, capturing the fact that equal-noise branches continue to support balanced cloning and effective purification. Overall, these results indicate that diversity gains are severely constrained when noise scales with system size, and that the proposed framework naturally selects between cloning, purification, and direct transmission based on the underlying channel symmetry and noise growth.

\begin{figure}
    \centering

    \begin{subfigure}{0.9\linewidth}
        \centering
        \includegraphics[width=\linewidth]{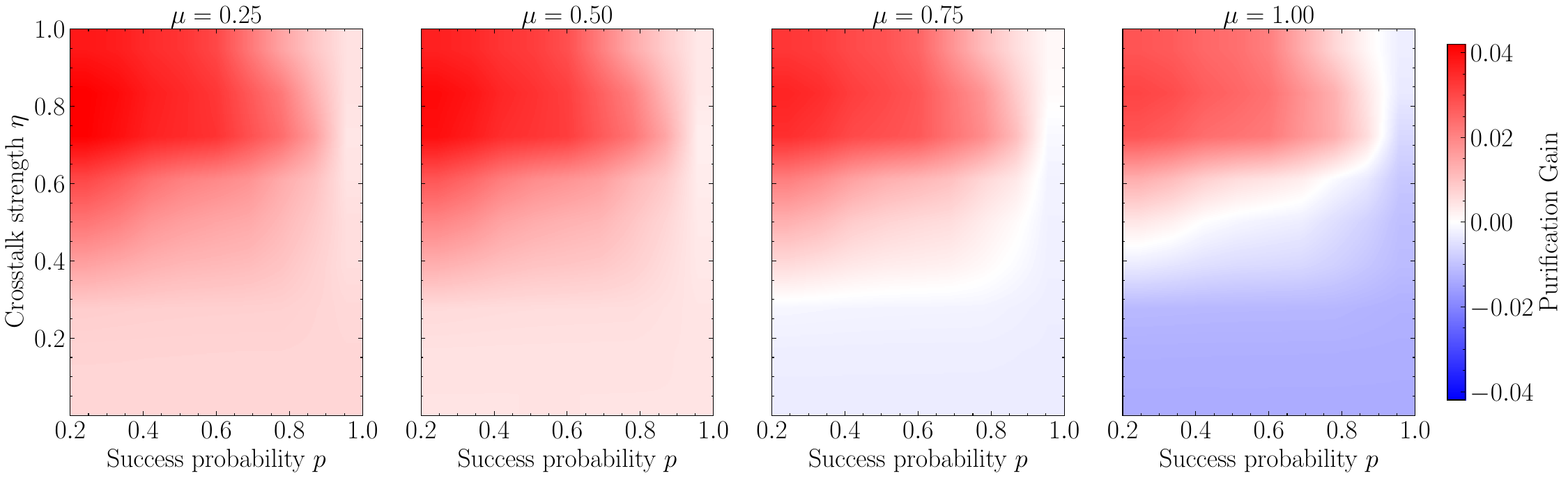}
        \caption{}
        \label{fig:gain_mean}
    \end{subfigure}

    \vspace{0.6em}

    \begin{subfigure}{0.9\linewidth}
        \centering
        \includegraphics[width=\linewidth]{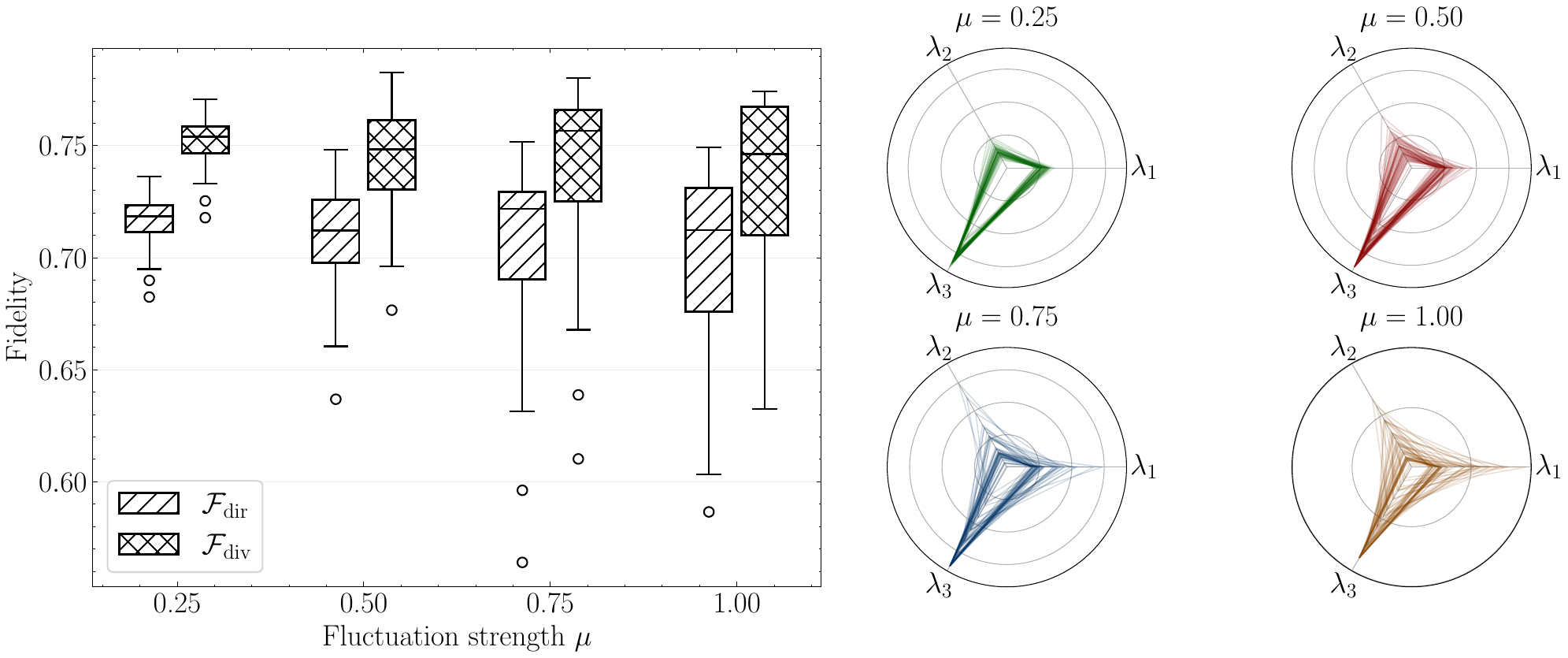}
        \caption{}
        \label{fig:gain_fluct}
    \end{subfigure}

    \caption{(a) Heatmap of the probabilistic purification gain as a function of success probability $p$ and crosstalk strength $\eta$ for a three–channel configuration ($M=3$) with total depolarization budget $Z=1.2$. The baseline fidelity at $p=1$ is obtained by deterministically purifying an ensemble of $50$ mean depolarization vectors $\boldsymbol{\lambda}^{(m)}$. For each $(p,\eta)$ pair, $50$ stochastic realizations per mean vector are generated to emulate random fluctuations around the average channel state. The purification gain $\mathcal{G}(p)$, defined as the deviation of $\mathcal{F}_{\mathrm{div}}(p)$ from the baseline fidelity $\mathcal{F}_{\mathrm{div}}(p=1)$, reveals how stochastic depolarization fluctuations influence the fidelity–success probability tradeoff under varying crosstalk conditions.
    (b) Realization–level statistics at a single operating point $\boldsymbol{\lambda}^{(0)}$, where we fix $p=0.8$, $\eta=0.8$ and examine fluctuations explicitly: the boxplots compare the distributions of $\mathcal{F}_{\mathrm{dir}}$ and $\mathcal{F}_{\mathrm{div}}$ across fluctuation strengths $\mu$, while the spider plots show the corresponding spread of the three–channel depolarization vectors $\boldsymbol{\lambda}$ for each $\mu$.}
    \label{fig:gain}
\end{figure}

\subsubsection*{Stochastic Depolarizing Channel}

In this setting, the depolarizing parameter associated with each branch fluctuates randomly around its mean value, producing a stochastic channel whose instantaneous noise realization is unknown to both transmitter and receiver. These fluctuations disrupt the deterministic ranking of the branches, making it difficult to predict whether asymmetric cloning or selective transmission remains advantageous on average. This behavior is illustrated in Fig.~\ref{fig:gain}. The mean gain heatmap in Fig.~\ref{fig:gain_mean} illustrates the
purification gain,
\begin{align}
    \mathcal{G}(p) = \mathcal{F}_{\mathrm{div}}(p) - \mathcal{F}_{\mathrm{div}}(1),
\end{align}
which measures the fidelity improvement achieved by allowing a
probabilistic ($p<1$) purification step over the deterministic baseline
decoder ($p=1$).  The results show that, within the
\blue{adaptive cloning–purification diversity scheme}, this gain remains
positive over a wide range of success probabilities and crosstalk
strengths, demonstrating that probabilistic purification continues to
provide a robust advantage even under substantial depolarization
fluctuations. The realization-level analysis in Fig.~\ref{fig:gain_fluct} further confirms that, even when instantaneous channel conditions deviate significantly from their mean, purification reliably mitigates these fluctuations, whereas direct transmission exhibits larger variance and degradation. These observations indicate that, under stochastic noise, the symmetric or near-symmetric allocation produced by purification offers an ergodically stable strategy, outperforming branch-ranking approaches that rely on instantaneous channel ordering.

\section*{Discussion}\label{disc}
In this work, we developed a numerical framework for analyzing \ac{QuMIMO} systems and for identifying principled guidelines for encoder design, channel modeling, and decoding strategies. Our results show that the cloning–purification architecture can effectively exploit quantum spatial diversity, particularly in regimes where crosstalk is strong and quantum information becomes highly mixed across branches. These insights establish a baseline for investigating more sophisticated channel models, encoding schemes, and recovery strategies within a unified simulation environment.
Regarding future directions, an immediate extension is to generalize the diversity concept to system models involving more than a single quantum input state. In such scenarios, the presence of crosstalk causes the input states to become coherently mixed, necessitating a unified diversity–multiplexing framework. Developing an appropriate codec that balances the fidelity of each communicated quantum state—or the average fidelity of a quantum stream across multiple channel uses—constitutes a central challenge in this setting. Another promising direction is to consider the transmission of a subsystem of an entangled state. In entanglement distribution, the entangled resource can be generated on demand, suggesting that multiplexing as many subsystems as possible may be advantageous, provided the received fidelity remains above the required threshold. In contrast, for quantum sensing applications, the transmitted subsystem encodes the sensed information within its phase component. Determining the optimal probing and \blue{measurement} strategy in such crosstalk-affected quantum sensing channels remains an open problem. Moreover, \ac{QEC}-inspired stabilizer codes represent a compelling complementary approach. While their primary purpose is to correct errors rather than enhance diversity, they may nonetheless provide meaningful improvements in end-to-end fidelity within high-dimensional \ac{QuMIMO} architectures, particularly as system size and noise complexity increase.

\section*{Methods}\label{Methods}
To establish our main \nameref{res} section, we outline the key techniques underlying the proposed \ac{QuMIMO} framework as shown in Fig.~\ref{fig:block}.

\subsection*{Encoding: Universal Asymmetric Quantum Cloning}

The transmitter encodes the input qubit state $\boldsymbol{\rho}$ using a 
universal asymmetric quantum cloning map 
$\mathcal{E}_M^{\boldsymbol{\gamma},\boldsymbol{t}}(\cdot)$ 
that produces $M$ approximate copies of the input.  
While universal cloning requires a physically realizable process acting on 
all possible inputs, state-dependent cloning can also employ virtual or 
Hermitian-preserving maps to achieve perfect fidelity for restricted 
state sets~\cite{BQY:25:PRA}. In the universal case considered here, the 
transformation is modeled as a \ac{CPTP} map
\begin{align}
   \mathcal{E}_M^{\boldsymbol{\gamma},\boldsymbol{t}}:
   \mathcal{B}(\mathbb{C}^2)
   \longrightarrow
   \mathcal{B}\!\left((\mathbb{C}^2)^{\otimes M}\right),
   \label{eq:cloning_map_def}
\end{align}
parameterized by an \emph{asymmetry vector}
$\boldsymbol{\gamma} = (\gamma_1,\ldots,\gamma_M)$
with non-negative components satisfying 
$\sum_{j=1}^{M}\gamma_j = 1$. Each element $\gamma_j$ specifies the relative fraction of quantum 
information allocated to the $j$-th clone. The reduced state of clone $k$ 
is isotropic and follows a depolarizing form,
\begin{align}
   \boldsymbol{\rho}_{\mathrm{c},k}
   = \Tr_{\{j\neq k\}}\!\left[
       \mathcal{E}_M^{\boldsymbol{\gamma},\boldsymbol{t}}(\boldsymbol{\rho})
     \right]
   = p_k\,\boldsymbol{\rho}
     + (1-p_k)\,\tfrac{\boldsymbol{I}}{2},
   \label{eq:clone_marginal}
\end{align}
where $p_k\!\in\![0,1]$ denotes the depolarizing strength.  
The corresponding single-clone fidelity is
\begin{align}
   F_k
   = \bra{\psi}\boldsymbol{\rho}_{\mathrm{c},k}\ket{\psi}
   = \tfrac{1+p_k}{2}, 
   \qquad
   \tfrac{1}{2} \le F_k \le 1.
   \label{eq:clone_fidelity}
\end{align}

\begin{definition}[Cloning isometry]
The asymmetric cloner is realized via a Stinespring isometry 
$V_{\boldsymbol{\gamma}}$ acting jointly on the input qubit and an 
environment,
\begin{align}
   V_{\boldsymbol{\gamma}}\ket{\psi}
   = \sum_{k=1}^{M}
      \beta_k\,\mathrm{Sym}_M
      \!\left(
         \ket{\psi}_k \!\otimes\!
         \bigotimes_{\substack{j=1 \\ j\neq k}}^{M}\!\ket{0}_j
      \right)
      \!\otimes\!\ket{e_k}_{\mathrm{env}},
   \label{eq:cloner_isometry}
\end{align}
where $\mathrm{Sym}_M$ projects onto the symmetric subspace of 
$(\mathbb{C}^2)^{\otimes M}$ and 
$\{\ket{e_k}_{\mathrm{env}}\}$ is an orthonormal environmental basis. The corresponding channel and its Choi operator are given by
\begin{align}
   \mathcal{E}_M^{\boldsymbol{\gamma}}(\boldsymbol{\rho})
   &= \Tr_{\mathrm{env}}\!\left[
       V_{\boldsymbol{\gamma}}\,\boldsymbol{\rho}\,
       V_{\boldsymbol{\gamma}}^{\dagger}
     \right],
     \label{eq:cloner_channel}\\[3pt]
   \boldsymbol{J}^{\mathcal{E}}(\boldsymbol{\gamma})
   &= \Tr_{\mathrm{env}}\!\left[
       (V_{\boldsymbol{\gamma}}\!\otimes\!I)
       \Phi_2
       (V_{\boldsymbol{\gamma}}^{\dagger}\!\otimes\!I)
     \right],
     \label{eq:cloner_choi}
\end{align}
where $\Phi_2 = \ketbra{\Phi_2}$ and 
$\ket{\Phi_2} = (\ket{00}+\ket{11})/\sqrt{2}$.
\end{definition}

\begin{figure}[t]
  \centering
  \begin{minipage}{0.9\textwidth} 
    \centering

    \begin{minipage}[b]{0.4\textwidth}
      \centering
      \includegraphics[width=\textwidth]{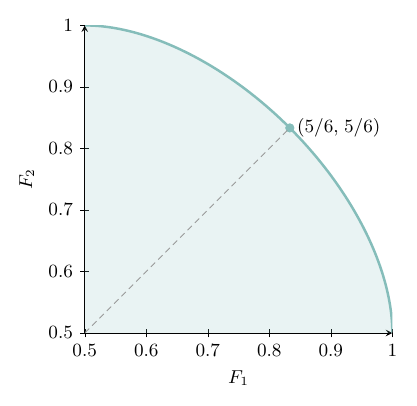}
      \subcaption*{(a) $M=2$}
    \end{minipage}
    \begin{minipage}[b]{0.4\textwidth}
      \centering
      \includegraphics[width=\textwidth]{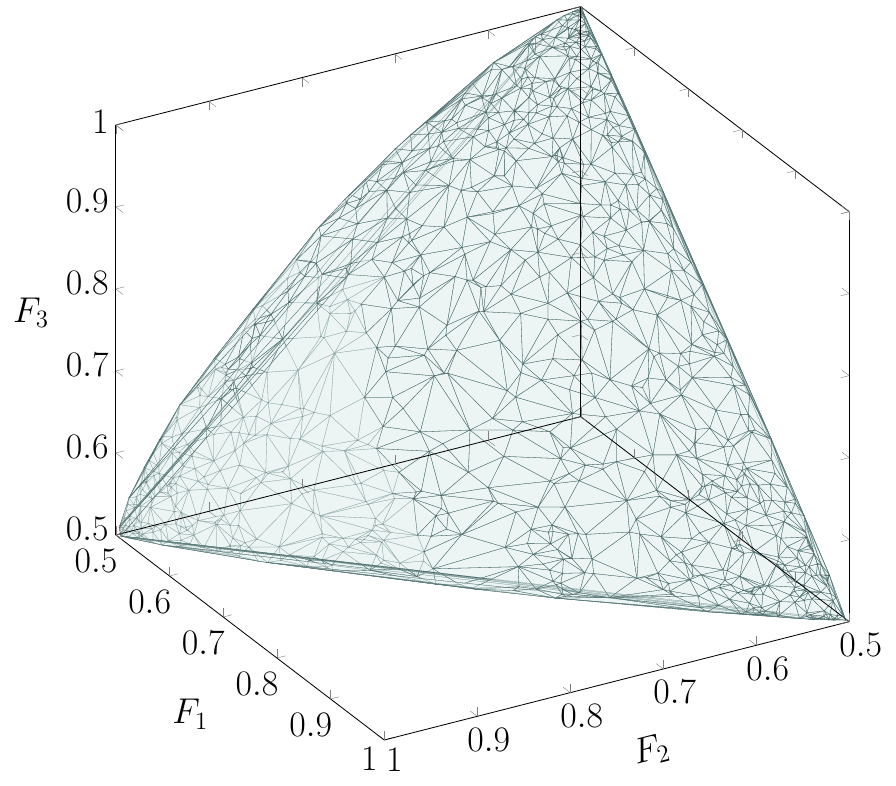}
      \subcaption*{(b) $M=3$}
    \end{minipage}

    \caption{Feasible fidelity trade-off regions of the universal asymmetric quantum cloner.
    (a) Achievable fidelity boundary for the $1{\rightarrow}2$ cloning scenario, illustrating how improving one clone’s fidelity necessarily degrades the other.
    (b) Convex-hull representation for $1{\rightarrow}3$ cloning, depicting the complete set of achievable fidelity combinations across three output channels.
    The lower bound of $0.5$ corresponds to the fidelity of a maximally mixed qubit, serving as the fundamental noise limit. 
    These trade-off surfaces characterize the intrinsic constraints of asymmetric cloning in the proposed quantum MIMO framework, where the clone fidelity imbalance reflects the resource distribution across  quantum channels.}
    \label{fig:cloning-tradeoff}
  \end{minipage}
\end{figure}

\begin{definition}[Fidelity boundary]
Tracing out all but the $k$-th clone yields a depolarizing marginal 
whose single-clone fidelity depends on the encoder asymmetry vector 
$\boldsymbol{\gamma}$ as
\begin{align}
   F_k(\boldsymbol{\gamma})
   = \frac{1}{3}
     + \frac{\big(2\,\beta_k(\boldsymbol{\gamma})
                  + \sum_{j=1}^{M}\beta_j(\boldsymbol{\gamma})\big)^{2}}{6},
   \label{eq:fidelity_gamma}
\end{align}
with $\boldsymbol{\beta}(\boldsymbol{\gamma})
 = (\beta_1(\boldsymbol{\gamma}),\ldots,\beta_M(\boldsymbol{\gamma}))$
determined by the normalized Perron–Frobenius eigenvector of the 
weight matrix
\begin{align}
   A(\boldsymbol{\gamma})
   = \boldsymbol{\alpha}\,\mathbf{1}^{\mathsf{T}}
     + \mathrm{diag}(\boldsymbol{\alpha}),
   \qquad
   \alpha_j
   = \frac{\gamma_j}{\sum_{k=1}^{M}\gamma_k},
   \label{eq:weight_matrix}
\end{align}
where $\mathbf{1}$ is the all-ones vector.

The eigenvector $\boldsymbol{u}(\boldsymbol{\gamma})$ associated with 
the largest (Perron) eigenvalue determines the physically realizable 
amplitude distribution, with normalized coefficients
\begin{align}
   \beta_j(\boldsymbol{\gamma})
   = \sqrt{
       \frac{2}{
          \big(\sum_{k=1}^{M} u_k(\boldsymbol{\gamma})\big)^{2} + 1
       }
     }\,u_j(\boldsymbol{\gamma}),
   \label{eq:beta_coeff}
\end{align}
ensuring correct normalization of the global Stinespring isometry.

The resulting fidelity vector 
$\boldsymbol{F}(\boldsymbol{\gamma})
 = (F_1,F_2,\ldots,F_M)$
specifies a point on the \emph{universal asymmetric cloning boundary}.  
The full set of achievable fidelity triples is the convex hull of 
$\boldsymbol{F}(\boldsymbol{\gamma})$ over all 
$\boldsymbol{\gamma}\in\Delta_M$, as illustrated in 
Fig.~\ref{fig:cloning-tradeoff} for $M{=}2,3$.
\end{definition}

In numerical simulations, the cloning operation is implemented through 
its Choi operator $\boldsymbol{J}^{\mathcal{E}}$, which satisfies
\begin{align}
   \Tr_{\mathrm{out}}\!\left[\boldsymbol{J}^{\mathcal{E}}\right]
   = \tfrac{I_d}{d},
   \label{eq:choi_tp_condition}
\end{align}
and acts on inputs via
\begin{align}
   \mathcal{E}_M^{\boldsymbol{\gamma},\boldsymbol{t}}(\boldsymbol{\rho})
   = d\,\Tr_{\mathrm{in}}\!\Big[
        \boldsymbol{J}^{\mathcal{E}}\,
        (I_{d^M}\!\otimes\!\boldsymbol{\rho})
      \Big].
   \label{eq:apply_cloner_sim}
\end{align}

For each Haar-random input state $\ket{\psi}$, the corresponding 
multi-clone output is
\begin{align}
   \boldsymbol{\rho}_{\mathrm{c}}
   = \mathcal{E}_M^{\boldsymbol{\gamma},\boldsymbol{t}}(\ketbra{\psi}),
   \label{eq:multi_clone_output}
\end{align}
whose local marginals $\boldsymbol{\rho}_{\mathrm{c},k}$ achieve the 
target fidelities $\{F_k\}$ and represent the operational asymmetry 
point of the cloning process.

\subsection*{Noise Sampling Under Average Channel Statistics}

To generate realistic QuMIMO noise under average channel statistics, we
treat the available information as a mean depolarization allocation
\begin{align}
  \boldsymbol{\lambda}^{(m)}
  = (\lambda_1^{(m)},\ldots,\lambda_N^{(m)}) \in [0,1]^N,
  \qquad 
  \sum_{i=1}^N \lambda_i^{(m)} = Z,
\end{align}
sampled from a Dirichlet distribution supported on the simplex
\begin{align}
  \mathcal{A}_Z
  = \{\boldsymbol{\lambda}\in[0,1]^N : \sum_{i=1}^N \lambda_i = Z\}.
\end{align}

To model deviations around this average channel description, we apply
multiplicative Gamma--Gamma perturbations following optical turbulence
models \cite{HAP:01:OE}, which describe irradiance fluctuations as the
product of two independent Gamma-distributed processes associated with
small- and large-scale scintillation. Each
instantaneous realization is
\begin{align}
  \mathbf{x}^{(m)}(\mu)
  = \Pi_{\mathcal{A}_Z}\!\left(
      \boldsymbol{\lambda}^{(m)} \odot \boldsymbol{\xi}(\mu)
    \right),
\end{align}
where $\odot$ denotes the Hadamard product and $\Pi_{\mathcal{A}_Z}$
projects back onto $\mathcal{A}_Z$ to preserve feasibility. The
fluctuation vector is defined as
\begin{align}
  \boldsymbol{\xi}(\mu)
  = (\xi_1(\mu),\ldots,\xi_N(\mu)),
  \qquad 
  \xi_i(\mu) = g_{1,i}\,g_{2,i},
\end{align}
with
\begin{align}
  g_{k,i} \sim \Gamma\!\bigl(c(\mu),1/c(\mu)\bigr),
\end{align}
chosen such that
\begin{align}
  \mathbb{E}[\xi_i(\mu)] = 1,
  \qquad
  \mathrm{Var}[\xi_i(\mu)] = \mu^2.
\end{align}
The shape parameter ensures that the targeted variance is
\begin{align}
  c(\mu)
  = \frac{1+\sqrt{1+\mu^2}}{\mu^2}.
\end{align}

This Gamma--Gamma stochastic formulation yields positive, unit-mean fluctuations
superimposed on the mean allocation, providing a controlled mechanism
for studying the sensitivity of the proposed \ac{QuMIMO} framework to
stochastic variations in depolarization levels.

\subsection*{Decoding: Probabilistic Purification}

Upon reception, Bob combines the received qubits using a probabilistic decoding map
$ \mathcal{D}_{K}^{\boldsymbol{r},p}(\cdot) $,
where $ \boldsymbol{r} $ denotes the $K$-element index vector specifying the selected subchannels
and $p$ represents the success probability of the purification.
A failed decoding attempt is assumed to yield no information 
about the transmitted state and is therefore modeled as a 
maximally mixed state 
$\boldsymbol{\omega} = I/2$.
The resulting average output state is
\begin{align}
   \hat{\boldsymbol{\rho}}
   = p\,\boldsymbol{\rho}_{\mathrm{p}} 
     + (1-p)\,\boldsymbol{\omega},
   \label{eq:average_output_state}
\end{align}
where $\boldsymbol{\rho}_{\mathrm{p}}$ denotes the successfully purified 
output. The overall end-to-end transformation of the system is thus
\begin{align}
   \hat{\boldsymbol{\rho}}
   =
   \mathcal{D}_{K}^{\boldsymbol{r},p}\!\Big(
   \mathcal{H}_{N}^{\eta,\boldsymbol{\lambda}, \delta}\!
   \big(\mathcal{E}_M^{\boldsymbol{\gamma},\boldsymbol{t}}(\boldsymbol{\rho})\big)\Big).
\end{align}

The decoder
$ \mathcal{D}_{K}^{\boldsymbol{r},p}(\cdot) $
is modeled as a completely positive trace–non-increasing (CPTNI) map
with Choi operator
$ \boldsymbol{J}^{\mathcal{D}}_{A\rightarrow B} $
acting from the $K$-qubit input space $A$ to the single-qubit output space $B$.
The optimal purification map maximizing the output fidelity can be obtained through the following \ac{SDP}~\cite{YCH:25:QST}:
\begin{align}
   &\max_{\boldsymbol{J}^{\mathcal{D}}_{A\rightarrow B}}
   \frac{1}{p}\,
   \Tr\!\left[\boldsymbol{J}^{\mathcal{D}}_{A\rightarrow B}
   \boldsymbol{Q}^{\mathrm{T}_A}\right]
   \label{eq:SDP_purification}\\
   \text{s.t.}\quad
   &\Tr\!\left[\boldsymbol{J}^{\mathcal{D}}_{A\rightarrow B}
   \boldsymbol{R}^{\mathrm{T}_A}\right] = p, \\
   &\boldsymbol{J}^{\mathcal{D}}_{A\rightarrow B} \ge 0,
   \qquad
   \Tr_B\!\left[\boldsymbol{J}^{\mathcal{D}}_{A\rightarrow B}\right]
   \le I_{2^K}.
\end{align}

\begin{definition}[Haar-averaged operators]
The operators $\boldsymbol{Q}$ and $\boldsymbol{R}$ are defined through
the Haar average over all pure input states $\boldsymbol{\rho}
= \ket{\psi}\!\bra{\psi}$ as
\begin{align}
    \boldsymbol{Q}
    &= \int \Lambda_A(\boldsymbol{\rho}) \otimes \boldsymbol{\rho}\, d\psi,\\
    \boldsymbol{R}
    &= \int \Lambda_A(\boldsymbol{\rho}) \otimes I\, d\psi,
    \label{eq:QR_state}
\end{align}
where $\Lambda_A$ denotes the composite encoder–channel map.
The operator $\boldsymbol{Q}$ characterizes the average input–output
correlation induced by $\Lambda_A$, while $\boldsymbol{R}$ provides the
corresponding Haar-averaged normalization term required for CPTNI
constraints within the purification SDP~\cite{ZL:04:PRA,YCH:25:QST}.
\end{definition}

Using the Choi–Jamiołkowski isomorphism and the Haar–twirling identity,
\begin{align}
   \int\boldsymbol{\rho}^{\otimes 2}\, d\psi
   = \frac{I + \mathbb{S}}{d(d+1)}
   = \frac{2}{d(d+1)}\,\Pi_{\mathrm{sym}},
   \label{eq:haar_twirling_identity}
\end{align}
the corresponding Choi expressions become
\begin{align}
   \boldsymbol{Q}
   &= \Tr_{\mathrm{env}}\!\left[
        (\boldsymbol{J}^{\mathcal{H}}\!\ast\!\boldsymbol{J}^{\mathcal{E}})
        (\boldsymbol{\rho}^{\mathsf{T}}\!\otimes I)
      \right],
   \label{eq:Q_choi}\\
   \boldsymbol{R}
   &= \Tr_{\mathrm{env}}\!\left[
        (\boldsymbol{J}^{\mathcal{H}}\!\ast\!\boldsymbol{J}^{\mathcal{E}})
        (I\!\otimes I)
      \right],
   \label{eq:R_choi}
\end{align}
where $\ast$ denotes the link product.

This formulation encapsulates the complete effect of the encoder–channel
cascade—parametrized by the cloning asymmetry vector $\boldsymbol{\gamma}$
and the CSI-dependent channel parameters $(\eta,\boldsymbol{\lambda},\delta)$. \blue{The Haar averaging in~\eqref{eq:average_fidelity} is encoded in
the operators $\boldsymbol{Q}$ and $\boldsymbol{R}$, which enter directly into the purification
\ac{SDP}~\eqref{eq:SDP_purification}, thereby linking the choice of decoding map and its
success probability $p$ to the resulting end-to-end average fidelity
$\overline{\mathcal{F}}$. In particular, the constraint
$\Tr(\boldsymbol{J}^{\mathcal{D}}\boldsymbol{R}^{\top_A})=p$ forces
the decoder to be deterministic \ac{CPTP} when $p=1$, but allows 
post-selection when $0<p<1$. Thus $\mathcal{D}^{(p=1)}$ can only act as
a valid trace-preserving decoder and effectively selects the best
available mode, whereas for $0<p<1$ the decoder may discard noisy
realizations and apply a stronger purification map on the accepted
outcomes. Consequently, reducing $p$ can strictly increase the
Haar-averaged fidelity in the \ac{QuMIMO} setting.}  When CSI is available,  
$ \boldsymbol{J}^{\mathcal{H}} $ is known
and the receiver constructs $\boldsymbol{Q}$ and $\boldsymbol{R}$ directly.
In the absence of CSI, the receiver employs a \textit{blind diversity model},
evaluating the integrals for a symmetric cloner and identity channel:
\begin{align}
    \mathcal{H}_{N}^{\eta,\pmb{\lambda}, \delta}(\cdot) &= \mathrm{Id}(\cdot), \\
    \gamma_k &= \tfrac{1}{M}\,, \qquad k=1,\ldots,M,
\end{align}
yielding the baseline $Q$ and $R$ for symmetric cloning without adaptation.
This blind approximation captures the ensemble-averaged behavior of the
QuMIMO link, allowing both perfect-CSI and non-CSI regimes to be treated
within the unified \ac{SDP} formulation~(\ref{eq:SDP_purification}).

\subsection*{Optimizing the Cloning Asymmetry}
Given the \ac{QuMIMO} channel 
$\mathcal{H}_{N}^{\eta,\boldsymbol{\lambda}, \delta}(\cdot)$,
the goal is to jointly optimize the encoder 
$\mathcal{E}_M^{\boldsymbol{\gamma},\boldsymbol{t}}(\cdot)$ 
and the probabilistic purification decoder 
$\mathcal{D}_{K}^{\boldsymbol{r},p}(\cdot)$ 
so as to maximize the end-to-end fidelity of the recovered quantum state.
The full optimization problem is formulated as
\begin{align}
   &\max_{\mathcal{E},\,\mathcal{D}}
   \; g\!\left(
      F\!\left(
      \mathcal{E}_M^{\boldsymbol{\gamma},\boldsymbol{t}},
      \mathcal{H}_{N}^{\eta,\boldsymbol{\lambda}, \delta},
      \mathcal{D}_{K}^{\boldsymbol{r},p}
      \right)
   \!\right),
   \label{eq:full_optimization}\\
   \text{s.t.}\quad
   &M,K \le N, \qquad
   \boldsymbol{\gamma} \in \Delta_M, \qquad
   \boldsymbol{t},\boldsymbol{r} \in \mathcal{P}_N^{M}, \qquad
   0 \le p \le 1,
\end{align}
where $g(\cdot)$ is an application-dependent utility function
(e.g., linear, exponential, or logarithmic in fidelity).
The set $\Delta_M$ denotes the $M$-dimensional probability simplex
governing the cloning asymmetry vector,
and $\mathcal{P}_N^{M}$ represents the admissible choices of transmitter
and receiver subchannel subsets drawn from the $N$ available modes. The fidelity $F(\cdot)$ depends on the CSI parameters 
$(\eta,\boldsymbol{\lambda},\delta)$ through the fixed channel 
$\mathcal{H}_{N}^{\eta,\boldsymbol{\lambda}, \delta}$,
and on the purification success probability $p$ through the decoder.
The optimization searches jointly over the encoder asymmetry 
$\boldsymbol{\gamma}$, the active mode selections $(\boldsymbol{t},\boldsymbol{r})$,
and the decoder map $\boldsymbol{J}^{\mathcal{D}}$,
using the Haar-averaged operators $\boldsymbol{Q}$ and $\boldsymbol{R}$
defined in Eqs.~(\ref{eq:Q_choi})–(\ref{eq:R_choi}).

\subsubsection*{Single-Channel Baseline}

In the low-noise, weak-crosstalk regime, the optimal configuration
degenerates to $M=K=1$. With CSI, the optimal transmit and receive
indices are chosen as
\begin{align}
    t^\star = \arg\min_i \lambda_i,
    \qquad
    r^\star = \arg\max_j F_{j\mid t^\star},
\end{align}
where $F_{j\mid t^\star}$ is the single-branch output fidelity in mode
$j$ when transmitted through $t^\star$ of the \ac{QuMIMO} channel.  
Thus,
\begin{align}
    \boldsymbol{t}=(t^\star), \qquad \boldsymbol{r}=(r^\star).
\end{align}

The encoder and decoder reduce to
\begin{align}
    \mathcal{E}_1(\rho)=\rho,
    \qquad
    \mathcal{D}_1^{p}(\rho)=p\rho+(1-p)\tfrac{I}{2},
\end{align}
representing direct transmission with single-mode probabilistic
purification. Without \ac{CSI}, ranking is not possible and the baseline is obtained by
choosing any fixed transmit–receive pair $(t_0,r_0)$, e.g., uniformly at
random.

\subsubsection*{Cloning–Purification Diversity Optimization}
In principle, both the encoder and decoder can be optimized
alternately via successive \acp{SDP}
to maximize the end-to-end channel fidelity~\cite{RW:05:PRL}.
This alternating (AO) strategy treats the encoder $\mathcal{E}$
and decoder $\mathcal{D}$ as separate convex blocks,
solving one \ac{SDP} for $\mathcal{D}$ with $\mathcal{E}$ fixed,
and another for $\mathcal{E}$ with $\mathcal{D}$ fixed, until convergence.
Although each subproblem remains convex,
the overall procedure requires multiple large-scale \acp{SDP}
and incurs exponential growth in the number of decision variables
with $N$ subchannels.
For qubit systems $(d{=}2)$,
the purification decoder
$\mathcal{D}\!:\!\mathbb{C}^{2^N}\!\!\to\!\mathbb{C}^{2}$
is represented by a Choi matrix of size
$2^{N+1}\times2^{N+1}$,
corresponding to approximately $4^{\,N+1}$ real decision variables.
Therefore, a joint encoder–decoder optimization scales as
$O(8^{\,N})$ in the total number of coupled variables,
rendering the full AO-based joint optimization
computationally intractable for realistic channel dimensions. In this work, the channel parameters
$(\eta,\boldsymbol{\lambda},\delta)$ are fixed
and the optimization is restricted to the
\emph{cloning asymmetry vector} $\boldsymbol{\gamma}$.
The effective operators
$\boldsymbol{Q}(\boldsymbol{\gamma})$ and
$\boldsymbol{R}(\boldsymbol{\gamma})$
constructed from
Eqs.~(\ref{eq:QR_state})–(\ref{eq:R_choi})
have dimensions
$2^{M+1}\!\times\!2^{M+1}$ for qubit systems,
since the universal cloning map
$\mathcal{E}_M^{\boldsymbol{\gamma},\boldsymbol{t}}$
acts from one qubit to $M$ output qubits.
Hence, the optimization complexity now scales
with the number of transmitted clones on $N$ subchannels. To achieve tractability,
we employ an eigenvalue-based spectral relaxation that eliminates the need for iterative optimization
and provides an analytical upper bound on the achievable fidelity \cite{L:15:MFME}.
Given the fixed channel realization and
the operators
$\boldsymbol{Q}(\boldsymbol{\gamma})$
and $\boldsymbol{R}(\boldsymbol{\gamma})$,
the decoder-optimal fidelity for a given asymmetry vector
$\boldsymbol{\gamma}$ can be expressed through the
\emph{generalized Rayleigh quotient}
\begin{align}
   F(\boldsymbol{\gamma})
   = \max_{\psi\neq 0}
     \frac{\psi^{\dagger}\boldsymbol{Q}(\boldsymbol{\gamma})\psi}
          {\psi^{\dagger}\boldsymbol{R}(\boldsymbol{\gamma})\psi},
   \label{eq:rayleigh_fidelity}
\end{align}
which is equivalently formulated as the generalized eigenvalue problem
\begin{align}
   \boldsymbol{Q}(\boldsymbol{\gamma})x
   = \nu\,\boldsymbol{R}(\boldsymbol{\gamma})x,
\end{align}
where $\nu$ denotes the eigenvalue.
Diagonalizing
$\boldsymbol{R}(\boldsymbol{\gamma})
 = U\,\mathrm{diag}(\rho)\,U^{\dagger}$
on its support yields
\begin{align}
   \boldsymbol{R}(\boldsymbol{\gamma})^{-\tfrac12}
   &= U\,\mathrm{diag}(\rho^{-\tfrac12})\,U^{\dagger},\\
   \boldsymbol{M}(\boldsymbol{\gamma})
   &= \boldsymbol{R}(\boldsymbol{\gamma})^{-\tfrac12}
      \boldsymbol{Q}(\boldsymbol{\gamma})
      \boldsymbol{R}(\boldsymbol{\gamma})^{-\tfrac12}.
\end{align}
The decoder-optimal fidelity is then given by the spectral radius
\begin{align}
   F(\boldsymbol{\gamma})
   = \nu_{\max}\!\Big(
       \tfrac12\!\big[\boldsymbol{M}(\boldsymbol{\gamma})
       + \boldsymbol{M}(\boldsymbol{\gamma})^{\dagger}\big]
     \Big),
\end{align}
and the optimal encoder asymmetry vector is obtained as
\begin{align}
   \boldsymbol{\gamma}^{\star}
   = \arg\max_{\boldsymbol{\gamma}\in\Delta_M\cap\mathcal{C}_M}
     F(\boldsymbol{\gamma}),
\end{align}
where $\Delta_M=\{\boldsymbol{\gamma}\!\ge\!0,\ \sum_k\gamma_k=1\}$
is the probability simplex, and
$\mathcal{C}_M$ denotes the universal asymmetric–cloning feasibility region. This eigenvalue formulation replaces an iterative sequence of \acp{SDP}
with a single spectral decomposition,
reducing computational complexity from exponential
($O(8^{N})$ in joint AO) to cubic
($O((2^{N+1})^{3})$) scaling per eigen-solve,
followed by a single convex \ac{SDP} used
to obtain the corresponding physical purification map.
The Rayleigh quotient, therefore, serves as an analytical surrogate
for optimizing the encoder asymmetry $\boldsymbol{\gamma}$,
and its rank-one certificate,
\begin{align}
   \boldsymbol{J}^{\mathcal{D}}_{\mathrm{ray}}(\boldsymbol{\gamma})
   = p\,\boldsymbol{R}(\boldsymbol{\gamma})^{-\tfrac12}
      \ketbra{v_{\max}}\,
      \boldsymbol{R}(\boldsymbol{\gamma})^{-\tfrac12},
\end{align}
where $\ket{v_{\max}}$ is the principal eigenvector of
$\boldsymbol{M}(\boldsymbol{\gamma})$ and $p\!\in[0,1]$,
provides a tight spectral upper bound on the achievable fidelity.
This certificate arises from a standard relaxation of
the purification \ac{SDP} in Eq.~(\ref{eq:SDP_purification}),
where the affine CPTNI constraint
$\Tr_B(\boldsymbol{J}^{\mathcal{D}})\!\le\! I$
is replaced by the scalar normalization
$\Tr[\boldsymbol{J}^{\mathcal{D}}\boldsymbol{R}(\boldsymbol{\gamma})]=p$
while maintaining positivity
$\boldsymbol{J}^{\mathcal{D}}\succeq 0$.
After determining $\boldsymbol{\gamma}^{\star}$,
the physical purification map is obtained
by solving the full CPTNI decoder \ac{SDP}
in Eq.~(\ref{eq:SDP_purification})
with the selected success probability $p$
and all operational constraints intact.


\section*{Data availability}
The datasets generated and/or analysed during the current study are available from the corresponding author upon reasonable request.

\section*{Code availability}
The source code used to support the findings of this study is available from the corresponding author upon reasonable request.

\backmatter

\clearpage

\bibliography{IEEEabbrv,sn-bibliography}   

\section*{Acknowledgements}


\section*{Funding}
The work of Shehbaz~Tariq and Symeon~Chatzinotas was supported by the project Lux4QCI (GA 101091508) funded by the Digital Europe Program, and the project LUQCIA Funded by the European Union – Next Generation EU, with the collaboration of the Department of Media, Connectivity and Digital Policy of the Luxembourgish Government in the framework of the RRF program.


\section*{Authors' contribution}
ST and SC contributed to the idea. ST and SC developed the theory and wrote the manuscript. SC improved the manuscript and supervised the research. All authors contributed to the analysis and discussion of the results and improved the manuscript. All authors read and approved the final manuscript.

\section*{Competing interests}
The authors declare no competing interests.


\end{document}